\documentclass[structabstract]{aa}  

\usepackage{graphicx}

\usepackage{auto-pst-pdf}
\usepackage{amsmath}
\usepackage{breqn}
\usepackage{float}

\begin{document}
%
  \title{The Earth as an extrasolar transiting planet - II: HARPS and UVES detection of water vapour, biogenic O$_2$, and O$_3$
   \thanks{Based on observations made with ESO Telescopes at the La Silla Paranal Observatory under programme ID 086.C-0448}}

\author{L. Arnold \inst{1}
          \and
          D. Ehrenreich\inst{2}
            \and
          A. Vidal-Madjar\inst{3}
          \and
          X. Dumusque\inst{4}
          \and
          C. Nitschelm\inst{5}
          \and
          R. R. Querel\inst{6,7}
           \and
          P. Hedelt\inst{8}
          \and
          J. Berthier\inst{9}
          \and
          C. Lovis\inst{2} 
          \and
          C. Moutou\inst{10,11}
           \and
          R. Ferlet\inst{3}
           \and
         D. Crooker\inst{12}
          }

\institute{Aix Marseille Universit\'e, CNRS, OHP (Observatoire de Haute Provence), Institut Pyth\'eas, (UMS 3470), 04870 Saint-Michel-l'Observatoire, France
              \email{Luc.Arnold@osupytheas.fr}
         \and
            Observatoire de Gen\`eve, Universit\'e de Gen\`eve, 51 ch. des Maillettes, 1290 Sauverny, Switzerland
          \and
         Institut d'Astrophysique de Paris, UMR7095 CNRS, Universit\'e Pierre \& Marie Curie, 98bis Boulevard Arago, 75014 Paris, France
         \and
         Harvard-Smithsonian Center for Astrophysics, 60 Garden Street, 02138 Cambridge, USA
         \and
         Unidad de Astronom{\'i}a, Facultad de Ciencias B\'asicas, Universidad de Antofagasta, Avenida Angamos 601, Antofagasta, Chile
         \and
         University of Chile, Department of Electrical Engineering, 2007 Tupper Avenue, Santiago, Chile
         \and
         National Institute of Water and Atmospheric Research (NIWA), Lauder, New Zealand
           \and
         Deutsches Zentrum f{\"u}r Luft  und Raumfahrt e.V. (DLR), Oberpfaffenhofen, 82234 Wessling, Germany
          \and
         Institut de M\'ecanique C\'eleste et de Calcul des \'Eph\'em\'erides, Observatoire de Paris, Avenue Denfert-Rochereau, 75014 Paris, France
          \and 
         Aix Marseille Universit\'e, CNRS, LAM (Laboratoire d'Astrophysique de Marseille) UMR 7326, 13388 Marseille, France 
         \and
         CFHT Corporation, 65-1238 Mamalahoa Hwy Kamuela, Hawaii 96743, USA
         \and
         Astronomy Department, Universidad de Chile, Casilla 36-D, Santiago de Chile, Chile
   }

   \date{Received November 13, 2013; accepted xxx}

 
  \abstract
   {The atmospheric composition of transiting exoplanets can be characterized during transit by spectroscopy. Detections of several chemical species have previously been reported in the atmosphere of gaseous giant exoplanets. For the transit of an Earth twin, models predict that biogenic oxygen (O$_2$) and ozone (O$_3$) atmospheric gases should be detectable, as well as water vapour (H$_2$O), a molecule linked to habitability as we know it on Earth.}
   {The aim is to measure the Earth radius versus wavelength $\lambda$ - or the atmosphere thickness $h(\lambda)$ - at the highest spectral resolution available to fully characterize the signature of Earth seen as a transiting  exoplanet.}
   {We present observations of the Moon eclipse of December 21, 2010. Seen from the Moon, the Earth eclipses the Sun and opens access to the Earth atmosphere transmission spectrum. We used two different ESO spectrographs (HARPS and UVES) to take penumbra and umbra high-resolution spectra from $ \approx 3100$ to 10400\AA. A change of the quantity of  water vapour above the telescope  compromised the quality of the UVES data. We corrected for this effect in the data processing. We analyzed the data by three different methods. The first method is based on the analysis of pairs of penumbra spectra. The second makes use of a single penumbra spectrum, and the third of all penumbra and umbra spectra.  }
 {Profiles $h(\lambda)$ are obtained with the three methods for both instruments. The first method gives the best result, in agreement with a model. The second method seems to be more sensitive to the Doppler shift of solar spectral lines with respect to the telluric lines. The third method makes use of umbra spectra which bias the result by increasing the overall negative slope of $h(\lambda)$. It can be corrected for this \textit{a posteriori} from results with the first method. The three methods clearly show the spectral signature of the Rayleigh scattering in the Earth atmosphere and the bands of H$_2$O, O$_2$, and  O$_3$. Sodium is detected. 
 Assuming no atmospheric perturbations, we show that the E-ELT is theoretically able to detect the $O_2$ A-band in 8~h of integration for an Earth twin at 10~pc.}
{Biogenic $O_2$, $O_3$, and water vapour are detected in Earth observed as a transiting planet, and, in principle, would be within reach of the E-ELT for an Earth twin at 10~pc.}


\titlerunning{Earth as a transiting exoplanet observed during the 2010 lunar eclipse}

\maketitle


\section{Introduction}
Observing planetary transits gives access to a set of properties of the  star+planet system (e.g. \cite{winn2009}). The relative size of the planet with respect to the star, given by the depth of the transit light-curve, is one of the most straightforward methods. Spectrally resolved observations have shown for gaseous giant exoplanets that transit the brightest stars that the transit depth varies with wavelength (the first to report this were \cite{charbonneau2002},  \cite{vidal-madjar2003},  \cite{vidal-madjar2004}) which reveal the signature of the planetary atmosphere. At certain wavelengths, the atmosphere indeed absorbs the stellar light and becomes opaque, which makes the planet look larger in front of the star and produces a deeper transit. The atmosphere of smaller terrestrial exoplanets close to or in the habitable zone (HZ) is in principle also accessible by this method (early works by \cite{schneider1992}, \cite{schneider1994}, and more recent models by \cite{ehrenreich2006}, \cite{kaltenegger2009}, \cite{ehrenreich2012}, \cite{garciamunoz2012}, \cite{snellen2013} and \cite{betremieux2013}). Recent simulations show that the next generation of instruments in the infrared or visible domains, that is the James Webb Space Telescope (JWST) and the European Extremely Large Telescope (E-ELT), should be able to offer a first access to characterizing terrestrial planets in the HZ and detecting biosignatures (here atmospheric O$_2$ and O$_3$) and molecules linked to habitability (\cite{rauer2011}; \cite{hedelt2013}; \cite{snellen2013}).

In this context, it is of prime interest to observe Earth as a transiting exoplanet.  Vidal-Madjar et al. (2010) observed the August 2008 partial lunar eclipse with the high-resolution SOPHIE \'echelle spectrograph at the 193~cm Haute-Provence telescope (\cite{perruchot2008}; \cite{bouchy2009}) and showed that the Earth atmosphere thickness versus wavelength can be derived from such observations. Earth observed from the Moon during a lunar eclipse indeed transits in front of the Sun and gives the possibility to study the Earth atmosphere as during a transit. These first observations also allowed one to identify molecular species in the atmosphere, that is water vapour (H$_2$O), a molecule linked to Earth habitabilty, biogenic oxygen (O$_2$), and ozone (O$_3$).  Ozone and oxygen are considered as reliable biosignatures for Earth-like planets (\cite{owen1980}; \cite{leger1993}; \cite{seager2013}) except for terrestrial planets outside the liquid-water habitable zone (\cite{schindler2000}). These species were also detected by Pall\'e et al. (2009) from umbra spectra taken during the same eclipse. However, Vidal-Madjar et al. (2010) pointed out that the Pall\'e et al. spectrum is not the transmission spectrum of Earth observed as a distant transiting planet. The umbra signal indeed results from photons that are highly scattered and refracted by the deep atmosphere. These photons strongly deviate from the star-planet direction and cannot be collected during the observation of a transit.  From penumbra observations, Vidal-Madjar et al. (2010) additionally detected   the thin high-altitude sodium Na I layer. The larger diameter of Earth in the blue due to increasing Rayleigh scattering at shorter wavelengths is also well visible, although it is partially blended with residuals from spectrograph orders correction. 

In this paper, we analyze new observations obtained during the lunar eclipse on Dec. 21, 2010 with the ESO spectrographs HARPS (from $ \approx 3800$ to 6900\AA, \cite{mayor2003}) and UVES (from $\approx 3200$ to 10400\AA, \cite{dekker2000}). 

We present first the principle of the observation and different reduction methods:  the original Vidal-Madjar et al. (2010) data reduction method (method~1), which makes use of a pair of penumbra spectra, and a second method (method~2), which requires only one single penumbra spectrum to recover the atmosphere thickness profile (Sect.~\ref{principle}). We also present a third alternative method to derive the thickness profile, which makes use of penumbra and umbra spectra. This third approach is biased by the refraction in the Earth atmosphere (Sect.~\ref{method3}, method~3), but can be corrected \textit{a posteriori} with the result of method~1. An observation log and more details about the spectrographs are given in Sect.~\ref{log}. We give details on the data reduction in Sect.~\ref{HARPS-data-processing} and Sect.~\ref{UVES-data-processing} for HARPS and UVES, respectively.  The altitude profiles versus wavelength obtained with the three different methods are detailed in Sect.~\ref{Results}, as well as simulations of the E-ELT observing a transiting Earth twin at 10~pc.

\section{Principle of the observation (method~1 and 2)}
\label{principle}
The principle of Vidal-Madjar et al. (2010) method is the following: For an observer on the Moon in the deep penumbra, the Sun appears above the Earth limb as a small crescent.  The light in the penumbra is a mix of direct solar light that did not interact with Earth with light that passed through the (mostly upper) Earth atmosphere. The nature of this mixed light in the penumbra is similar to what is observed during a transit of an exoplanet, here Earth in front of the Sun. Nevertheless, as seen from the Moon, Earth is larger than the Sun, and the Earth radius versus wavelength cannot be reconstructed from the transit depth at all wavelengths, as done with transiting exoplanets.  The idea thus consists of observing two different positions of the Sun above the Earth limb (during ingress or egress) - that is, of recording Moon spectra at two different depths in the penumbra (method~1).
For these two different positions on the Moon, the solar crescent above the Earth limb as well as the arc of the Earth atmosphere intersecting the solar disk, are different. We have shown (\cite{vidal-madjar2010}, Eq.~14 therein)  that $h$, the thickness profile of the atmosphere, or the altitude under which the atmosphere can be considered as opaque, can be written

\begin{multline}
h (\lambda) = \left( \frac{E_A (\lambda)}{E_A (\lambda_0)} \times \frac{F_A
(\lambda_0)}{F_A(\lambda)} - \frac{E_B (\lambda)}{E_B(\lambda_0)} \times \frac{F_B (\lambda_0)}{F_B (\lambda)} \right) \\   \div  \left( \frac{L_B}{S_B} - \frac{L_A}{S_A} \right),
\label{ratioAB}
\end{multline}
where $E$ and $F$ are the fluxes during the eclipse and outside the penumbra (almost full-Moon just before or after the eclipse), respectively. Subscripts $_A$ and $_B$ stand for the two different positions in the penumbra. The value $\lambda_0$ is an arbitrary wavelength for which we define $h(\lambda_0)=0$, and the symbols $S$ and $L$ represent geometrical parameters equal to the surface of the Sun crescent above the Earth limb and the length of the arc of the Earth atmosphere intersecting the solar disk for $h(\lambda_0)$, respectively (see Fig. 4 in Vidal-Madjar et al. 2010). Eq.~\ref{ratioAB}  shows that $h(\lambda)$ can thus be written as a function of differences between the measured fluxes divided by a factor of geometrical values. The numerical values for $S$ are recovered from the measured fluxes too, $E (\lambda_0)$ and  $F (\lambda_0)$,  following Eq.~11 of Vidal-Madjar et al. (2010),
\begin{equation}
  E (\lambda_0) =  F (\lambda_0) \times  \frac{S}{S_\odot},
  \label{S}
\end{equation}
where $S_\odot$ is the surface of the solar disk. At last, once the surfaces S are known, the lengths $L$ can be calculated quite simply from geometrical considerations knowing the apparent diameters of Earth and Sun, so $h(\lambda)$ can finally be calculated.

Eq.~\ref{ratioAB} above is obtained from the differences between Eq.~12 and 13 in Vidal-Madjar et al. (2010), each being written for one position on the Moon. These equations can be rewritten, for position A for example,
\begin{equation}
  h (\lambda)= \left( 1 - \frac{E_A (\lambda)}{E_A (\lambda_0)} \times \frac{F_A
  (\lambda_0)}{F_A (\lambda)} \right)     \times  \frac{S_A}{L_A} .
  \label{ratioA}
\end{equation}
We note that $h(\lambda)$ is fully calculable from Eq.~\ref{ratioA} above, thus with only one penumbra spectrum (method~2). Nevertheless, because $E$ and $F$ are recorded at different times in the night, the barycentric Earth radial velocity (BERV) leads to a different Doppler shift between the solar and telluric lines in $E$ and $F$ spectra. Since we are interested in the signature of the Earth atmosphere, all spectra are re-aligned on the telluric lines, which slightly shifts the solar lines between $E$ and $F$. Moreover, the solar lines integrated over the solar crescent seen from the Moon during the partial eclipse are also Doppler-shifted with respect to the lines of the full (uneclipsed) solar disk. The solar lines thus do not cancel out correctly in the ratio $\frac{E_A (\lambda)}{E_A (\lambda_0)} /  \frac{F_A(\lambda)}{F_A (\lambda_0)}$, and the obtained $h(\lambda)$ profile is noisier, although the overall shape is correct. With the difference in the left part of Eq.~\ref{ratioAB}, the solar lines cancel out much better and the $h(\lambda)$ profile is cleaner. This is discussed in more details in Sect.~\ref{result-method2}.

\section{An alternative method that approximates $h(\lambda)$ (method~3)}
\label{method3}
This section describes another method, different from that of Vidal-Madjar et al. (2010), to derive the thickness profile of Earth atmosphere. In a lunar eclipse, the radius of Earth is given by the radius of the umbra. 
The umbra radius can be calculated geometrically for a solid sphere that casts its shadow to the Moon, but in practice, the Earth atmosphere that scatters light along Earth limb, lead to a blurred edge between umbra and penumbra on the Moon surface. The radiance profile across the umbra edge has been measured in the past by several observers  (especially by Dubois from 1950 to 1980, \cite{danjon1925}; \cite{combaz1950}; \cite{reaves1952}; \cite{walker1957}). The irradiance of the eclipsed Moon as well as the umbra edge have also been modelized (\cite{link1969}; \cite{vollmer2008}; \cite{garciamunoz2011}) by an analysis of the light propagation from the Sun to the Moon through the Earth atmosphere. The observed magnitude versus position with respect to the umbra center looks approximately like a hyperbolic function (or sigmoid). The observations mentioned above indicate that magnitude typically drops by about 8 magnitudes from penumbra to umbra, the steepest region being typically 2~arcmin wide (\cite{dubois1954}). Unfortunately, the spatial resolution of published radial photometry across the umbra edge is only about 1~arcmin, and not sufficient to reveal a dependance of the umbra radius with wavelength. To fix the order of magnitude, with an Earth umbra radius of $ \approx 43$~arcmin, a variation of 10~km in the Earth radius (of  6371~km) translates into a change of the umbra radius of 4~arcsec.

\begin{figure}[h]
   \centering
   \includegraphics[width=9cm]{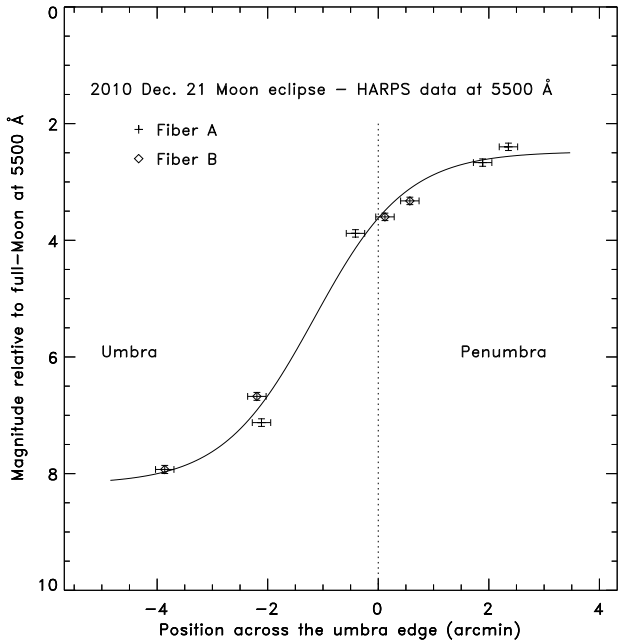}
   \caption{Measured magnitude across the umbra edge with HARPS at 5500\AA.  Crosses and diamonds stand for A and B fiber, respectively. The horizontal error bars are $2\sigma=20$~arcsec long and represent the typical pointing error of the telescope. Vertically, the $2\sigma$ error bars are the magnitude errors evaluated from the S/N ratio of the spectra at that wavelength and the evaluated error in the local Moon albedo. The solid line is the model from Eq.~\ref{eq:model} fitted through the data. The vertical dotted line marks our definition of the umbra edge. See Sect.~\ref{sect:result_method3} for details.}
    \label{fig:umbrapenumbraHARPS}
\end{figure}

With ESO's HARPS and UVES spectrographs, we recorded spectra from deep penumbra to umbra over $\approx 7$~arcmin with HARPS and $\approx 19$~arcmin with UVES. Because the telescope tracked neither the star nor the Moon but the Earth umbra projected to the Moon, we were able to  integrate the signal at fixed distances from the umbra edge, while the Moon itself slowly shifted in front of the spectrographs during the exposures. To compare the flux in these spectra taken at different positions in the umbra or in the penumbra, we calculate their corresponding magnitude $-2.5\ \log_{10}(E/F)$ relative to the full-Moon spectra recorded before the eclipse. Then we plot for each wavelength the magnitudes versus their distance from the umbra center, as Dubois did. But while Dubois observed at three or four different wavelengths through broadband blue, green, yellow, and red Wratten filters, HARPS and UVES allow to do this now at a much higher spectral resolution. The points follow the same overall shape as Dubois' profiles (Fig.~\ref{fig:umbrapenumbraHARPS}). We fit a hyperbolic function through these points 
\begin{equation}
 \zeta(r) = a_0+a_1\tanh[  a_3(r-a_2)],
  \label{eq:model}
\end{equation}
where $\zeta$ is the model giving the magnitude in the radial direction through the umbra edge, $r$ the distance to the umbra center, and $a_i$ the parameters defining the hyperbolic function. The function $\zeta(r)$ varies between $a_0-a_1$ and $a_0+a_1$.
We may arbitrarily define the abscissa $a_2$ of the inflection point (at ordinate $a_0$) of this function as a measurement of the umbra radius, and thus of the Earth radius at the observing wavelength. 

Nevertheless, Fig.~\ref{fig:umbrapenumbraHARPS} shows that the inflection point corresponds to a zone about 3 magnitudes ($\approx15\times$) fainter than shallower penumbra. But in a real transit, only the beams of light that interact with the top of the atmosphere, and thus only weakly absorbed, leave a signature in the transit. If a beam goes deeper to the atmosphere, it is significantly dimmed and scattered, and finally absent from the transit signature of the planet. 
Therefore the informations from the 3~mag fainter light we obtain at the inflection point are not those we obtain during a real transit. The inflection point is thus not the best definition of the edge of the Earth's umbra on the Moon, at least in the context of our Earth-transit analysis. It is more relevant to define a point slightly more in the brighter penumbra for the edge of Earth's umbra, as shown by the vertical dotted line in Fig.~\ref{fig:umbrapenumbraHARPS}, and to build up the Earth radius versus wavelength from this point instead of form the inflection point. More details are given in Sect.~\ref{sect:result_method3}.

We understand that this method makes use of information that is not available in an exoplanet transit: indeed it uses information from spectra of the umbra to build the function $\zeta$, but, as mentioned in the introduction, the photons from the umbra result from highly angularly scattered or refracted beams, which are not available in the observation of an exoplanet transit. Moreover, the umbra photons, especially the red ones, have been refracted by the deepest Earth atmospheric layers, and the Earth shadow is thus shrunk by refraction at these wavelengths, leading to a smaller Earth radius in the red. In other words, the gradient in the altitude profiles measured from blue to red wavelengths will be steeper due to refraction. But we show in Sect.~\ref{sect:result_method3} that the altitude profiles $h(\lambda)$ obtained by this method can be calibrated (scaled) to the profiles obtained with method~1.

\section{Observing log}
\label{log}
HARPS provides spectra with a spectral resolution of 115000 and covers the visible range (Tab.~\ref{tab:instr}, \cite{mayor2003}).
UVES was used in dichroic-1 and dichroic-2 modes (DIC1 and DIC2). In each of these modes, both BLUE and RED arms of the spectrograph are used simultaneously, and with standard settings, two exposures (one in each mode) give a full [3100-10400]\AA\  spectrum (Tab.~\ref{tab:instr}, \cite{dekker2000}). The slit width was set to 0.3~arcsec, which gives a spectral resolution of $\approx120000$.


\begin{table}[h]
  \caption{HARPS and UVES spectral ranges. Gaps without data are present in the final spectra due to small gaps between detectors. 
}
  \label{tab:instr}
  \begin{center}
    \leavevmode
    \begin{tabular}{lll} \hline \hline              
  Instrument      	       & range (\AA)     & gap (\AA) \\  \hline 
  HARPS           	& 3782-6912     &    5256-5337.5  	 \\ 
  UVES DIC1 BLUE   & 3055-3885	 & 	  \\
  UVES DIC1 RED     & 4788-6806	&  5750-5821	  \\
  UVES DIC2 BLUE   & 3751-4999	 & 	  \\
  UVES DIC2 RED   	&  6712-10426	&  8544-8646  \\  \hline
    \end{tabular}
  \end{center}
\end{table}

Tables \ref{tab:HARPS_ref} and \ref{tab:HARPS_fiber_pos} give information on the observations made from La Silla with HARPS. The position with respect to the umbra edge was measured with Guide-9\footnote{http:/www.projectpluto.com/}.  Tables \ref{tab:UVES_ref} and \ref{tab:UVES_slit_pos} give the same information but for UVES at Paranal. With UVES, the telescope was not positioned as expected on the Moon during the eclipse and the expected positions are indicated in parentheses in Tab.~\ref{tab:UVES_slit_pos}.

\begin{table}[h]
  \caption{Series of   $n$  Moon reference spectra taken with HARPS before the eclipse. }
  \label{tab:HARPS_ref}
  \begin{center}
    \leavevmode
    \begin{tabular}{lll} \hline \hline              
  start      			&  exposure      & airmass      \\  
  h:min:ss UT           	& ( $n\times$s)           	&                      \\  \hline 
  02:18:26   		& 4$\times$30	  	&        2.15-2.12              \\
  04:41:20   		&  5$\times$45		&        1.70             \\  \hline
    \end{tabular}
  \end{center}
\end{table}

\begin{table}[h]
  \caption{HARPS exposures and fiber positions in the penumbra (negative values) and umbra (positive values) with respect to the umbra edge given by Guide-9. }
  \label{tab:HARPS_fiber_pos}
  \begin{center}
    \leavevmode
    \begin{tabular}{llll} \hline \hline              
  start  &  exposure & Fib. A     & Fib. B    \\  
  h:min:ss UT           & (s)  &       (arcsec)       &(arcsec)              \\  \hline 
  06:43:32  & 	         600			& -66      &       41      \\
  06:57:28 & 		450			& -204    &     -98                	 \\
  07:06:35 & 		300			& -232    &     -125                \\
  07:13:07 & 		900			& 36        &      141                \\  \hline
    \end{tabular}
  \end{center}
\end{table}

\begin{table}[h]
  \caption{Series of  $3$ Moon reference spectra taken with UVES before the eclipse. }
  \label{tab:UVES_ref}
  \begin{center}
    \leavevmode
    \begin{tabular}{llll} \hline \hline              
  start &  exposure & mode and arm     & mean airmass    \\  
  h:min:ss UT           & ($3\times$s)  &              &              \\  \hline 
  02:04:19  & 	         3$\times$4			& DIC2 RED   &     2.01     \\
  02:04:23  & 	         3$\times$10			& DIC2 BLUE   &    2.01    \\
  02:18:49  & 	         3$\times$4			& DIC1 RED   &     1.90     \\
  02:18:56  & 	         3$\times$40			& DIC1 BLUE   &    1.90    \\

  05:07:25  & 	         3$\times$4			& DIC2 RED   &     1.56     \\
  05:07:29  & 	         3$\times$10			& DIC2 BLUE   &    1.56    \\
  05:19:26  & 	         3$\times$4			& DIC1 RED   &     1.57    \\
  05:19:30  & 	         3$\times$30			& DIC1 BLUE   &    1.58    \\
  
 \hline
   \end{tabular}
  \end{center}
\end{table}

\begin{table}[h]
  \caption{UVES exposures or series of $n$ exposures, and measured slit positions in the penumbra (negative values) and umbra (positive values) with respect to the umbra edge given by Guide-9. The expected slit positions are given in parentheses in the last column.}
  \label{tab:UVES_slit_pos}
  \begin{center}
    \leavevmode
    \begin{tabular}{llll} \hline \hline              
  start  &  exposure & mode \& arm    & slit position    \\  
  h:min:ss UT           & ($n\times$s)  &              &(arcsec)              \\  \hline 
  06:45:19  & 	         2$\times$40	& DIC1 RED   &     -210 (-90)     \\
  06:45:23  & 	         180			& DIC1 BLUE   &     -210 (-90)     \\
  06:51:07 & 		3$\times$20	& DIC1 RED   &    -233 (-180)               	 \\
  06:51:11 & 		180			& DIC1 BLUE    &    -233    (-180)            	 \\
 06:58:01 & 		600			& DIC1 RED   &    577  (360)              	 \\
  06:58:05 & 		600			& DIC1 BLUE   &    577   (360)             	 \\
  07:15:08 & 		2$\times$60	& DIC2 RED  &     -220   (-90)               \\
  07:15:12 & 		90			& DIC2 BLUE   &     -220    (-90)                    \\
  07:19:35 & 		2$\times$25	& DIC2  RED  &      -229   (-180)             \\  
  07:19:39 & 		60			& DIC2  BLUE  &      -229      (-180)              \\  
  07:22:30 & 		20			 & DIC2 RED   &     -147   (360)             \\
   07:22:34 & 		20			& DIC2  BLUE  &     -147       (360)              \\ 
  07:35:10 & 		600			& DIC2  RED  &    936  (360)              	 \\
  07:35:14 & 		600			& DIC2  BLUE  &    936   (360)         	 \\ 
   \hline
    \end{tabular}
  \end{center}
\end{table}

\begin{figure}[h]
   \centering
   \includegraphics[width=9cm]{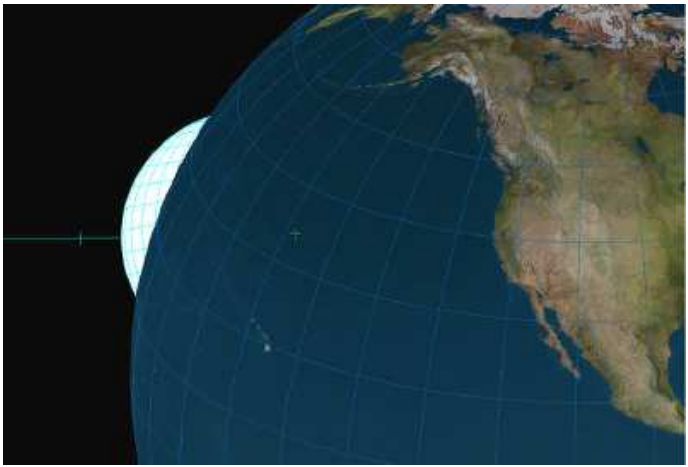}
   \caption{Earth and Sun seen from the Moon deep penumbra where HARPS fiber A was positioned at 07:06:35UT, at the beginning of the 300~s exposure. Image with Guide-9.}
    \label{Earth}
\end{figure}

As an example, we show in Fig.~\ref{Earth} the Sun seen from where HARPS fiber-A was positioned on the Moon at 07:06:35UT, at the beginning of the 300~s exposure (Tab. \ref{tab:HARPS_fiber_pos}). The Sun is setting in the Pacific Ocean about 2000~km southeast of Japan, at coordinates $\approx156^\circ$~E and $\approx23^\circ$~N. The daytime cloud fraction for 2010 Dec. 21 given by Aqua/MODIS\footnote{http://modis.gsfc.nasa.gov/} for these coordinates is $\ge\ \approx0.7$. 

\section{HARPS data processing}
\label{HARPS-data-processing}

The data processing makes use of \textsl{s1d} reduced data output files directly from the HARPS data reduction software (DRS). The HARPS instrumental response is calibrated with a high signal-to-noise ratio (S/N) G2V star spectrum.

\subsection{Correction for the Moon reflectivity}
\begin{figure}[h]
   \centering
   \includegraphics[width=9cm]{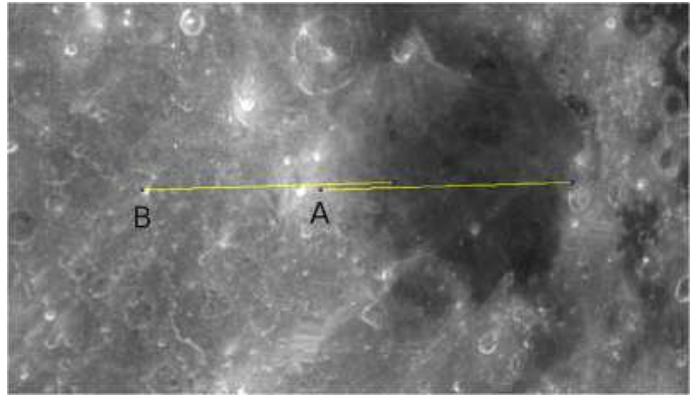}
   \caption{Example of a path of the HARPS fibers across the reconstructed Moon image, during the 300~s exposure in the penumbra. The darkening from the penumbra and umbra are not superimposed on the image here, since we are only interested in recovering the moon reflectivity. The A and B fibers remained  at 232 and 125~arcsec  from the umbra edge, while the fibers path across the Moon was $\approx164$~arcsec long (length of the yellow lines). The mean reflectivity is measured along the path, taking into account a 10~arcsec pointing error in both RA and DEC directions. The terrain reflectivity fluctuates by $\sigma = 21\%$ along the path for Fiber A and $\sigma = 15\%$ for Fiber B in this highly contrasted lunar region. The error on the mean reflectivity due to pointing uncertainty is estimated to be about  $\sigma = 4\%$. The figure shows that while Fiber A went through the dark region of Mare Nectaris, Fiber B remained on a region measured $39\%$ brighter between Catharina and Cyrillus craters, pointing out the relevance of correcting the flux for lunar local reflectivity.}
    \label{fig:slit_path_HARPS}
\end{figure}

To accurately maintain a given position of the spectrograph with respect to the umbra edge during the exposures, we needed to track the umbra, not the Moon itself. The Moon was thus moving in front of the spectrograph pair of fibers at a speed of about 0.55~arcsec/s. The shift of the Moon during a given exposure contributed to smooth out any relative spectral reflectance differences between the lunar terrains that drifted in front of the spectrograph. But different terrains with different albedos were seen by each fiber during the different exposures. To correct each spectrum for the flux variations induced by these albedo variations,  we measure the mean lunar reflectivity along the path of the slit across the Moon surface during the exposures \textsl{a posteriori}. The path is reconstructed from the date and telescope coordinates available in the metadata. We build a picture of the Moon as seen during the eclipse from La Silla at a resolution of 1~arcsec/pixel: the reconstructed scene is an orthographic projection of a Clementine albedo map\footnote{http://www.lpi.usra.edu/lunar/tools/clementine/} at 750~nm (\cite{eliason1999}) and takes into account the correct libration angles, the correct phase angle (close of zero degree), and a Lambert-Lunar limb-darkening law (\cite{mcewen1996}). The mean lunar reflectivity is then measured  with Audela-2.0\footnote{http://www.audela.org/}  along the path of each fiber across the lunar surface as shown in Fig.~\ref{fig:slit_path_HARPS}.  

\subsection{Correction for the signature of the atmosphere above the telescope}
To correct our eclipse spectra for the signatures of the atmosphere above the telescope, we need a reference transmission spectra for the atmosphere $T(\lambda)$. To derive $T(\lambda)$, we proceed as
in Vidal-Madjar et al. (2010) by calculating the ratio of two full-Moon spectra, $F_1$ and $F_2$,  taken before the eclipse at different airmass $AM_1$ and $AM_2< AM_1$. The ratio
can be scaled to AM=1 following 
\begin{equation}
     T(\lambda) =  {\bigg[  \frac{F_1^{AM_1} (\lambda)} {F_2^{AM_2} (\lambda)} \bigg]}^{(1/(AM_1-AM_2)}.
     \label{eq:am}
\end{equation}

\begin{figure}[h]
   \centering
   \includegraphics[width=9cm]{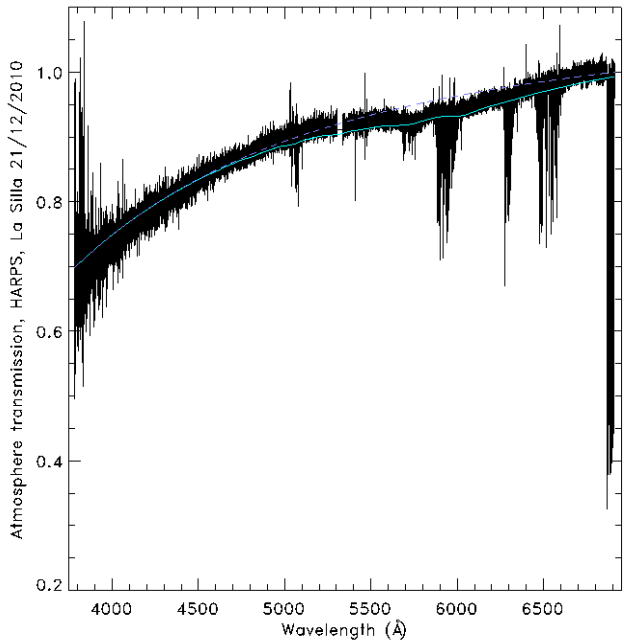}
   \caption{La Silla atmosphere transmission function $T(\lambda)$ for AM=1 as evaluated
   from full-Moon spectra taken with HARPS just before the eclipse (Tab.~\ref{tab:HARPS_ref}). The spectrum is normalized at 6866\AA.
   To verify this experimental $T(\lambda)$ spectrum, a transmission fit based on the model of 
   Hayes and Latham (1975) for the Rayleigh and aerosol components normalized at 6866\AA\ is displayed (blue dashed line). We also add an ozone transmission spectrum to the model (light-blue solid line) that correctly fits the observed wide Chappuis band. The best least-squares fit for Rayleigh and ozone over La Silla that night is obtained for an airmass of AM=0.77 for ozone and Rayleigh, and an optical depth of 0.181 at 5320\AA\ for the aerosols, between 4000 and 6866\AA. The fit for ozone is made only where O$_2$ and H$_2$O absorption bands are absent.  The O$_2$ $\gamma$ and B absorption bands are clearly seen, around 6300 and 6900\AA\ respectively, the other deep bands are due to H$_2$O. The small gap in HARPS data is visible between 5300 and 5340\AA\ (Tab.~\ref{tab:instr}).}
   \label{transmission}
\end{figure}

We calculate $T (\lambda)$ with $AM_1=2.2$ and $AM_2=1.7$ (Tab.~\ref{tab:HARPS_ref}). Since we are interested only in the accurate measurement of the telluric lines in $T(\lambda)$, the two spectra in Eq.~\ref{eq:am} are corrected for the BERV to properly align the telluric lines of both spectra (the solar lines become slightly misaligned).
Fig.~\ref{transmission} shows the obtained average reference transmission $T(\lambda)$ that
we now use in the following data analysis of HARPS data to re-evaluate all eclipse spectra as if they were collected from outside the atmosphere.

\section{UVES data processing}
\label{UVES-data-processing}
The UVES instrumental response is calibrated with the reference star HD80170 (K5 III-IV, T$_{eff}$=4537~K, $\log(g)=2.1$, \cite{randich1999}).  

\subsection{Correction for the Moon reflectivity}
\begin{figure}[h]
   \centering
   \includegraphics[width=9cm]{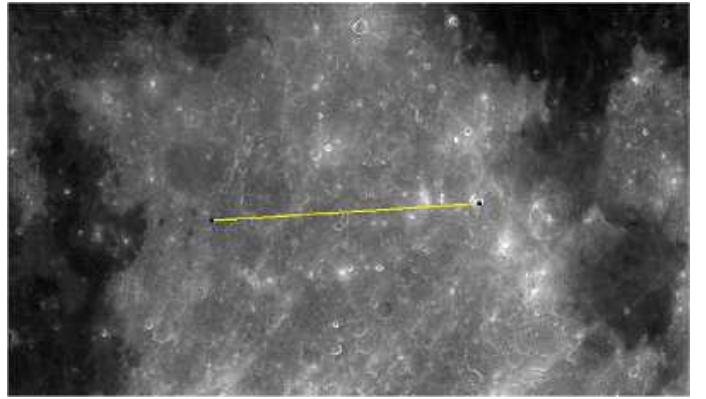}
   \caption{Example of a path of the UVES slit across the reconstructed Moon image during the 600~s exposure in the umbra. The slit, 10~arcsec long and perpendicular to the path, remains at $\approx936$~arcsec  from the umbra edge, while the slit path over the Moon was $\approx330$~arcsec long (length of the yellow line). The mean reflectivity is measured along the path, taking into account a 10~arcsec pointing error in both RA and DEC directions. The terrain reflectivity fluctuates by $\sigma = 13\%$ along the path in this bright lunar region, and the error on the mean reflectivity due to pointing uncertainty is estimated to be about  $\sigma = 4\%$. }
    \label{fig:slit_path_UVES}
\end{figure}

To measure the mean lunar reflectivity along the path of the UVES slit across the Moon surface, we build a picture of the Moon as seen during the eclipse from Paranal, as we did for the HARPS observations. An example of the slit path is shown in Fig.~\ref{fig:slit_path_UVES}.  

\subsection{Correction for the signature of the atmosphere above the telescope}
\label{section_UVES_trans}
\begin{figure}[h]
   \centering
   \includegraphics[width=9cm]{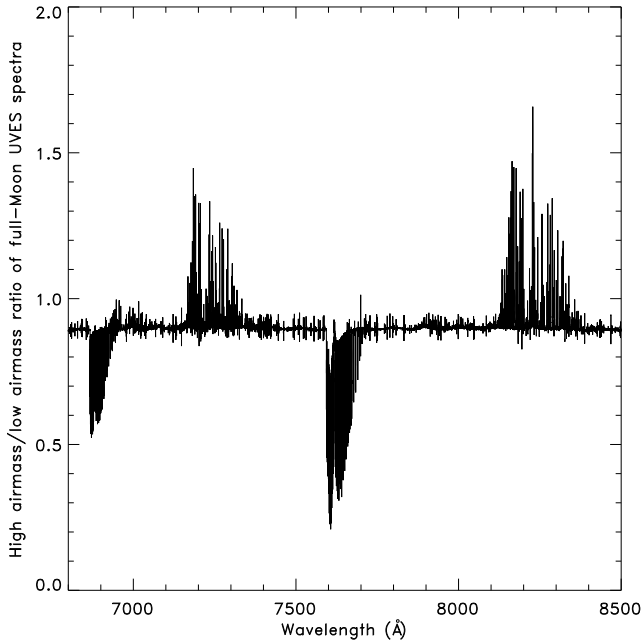}
   \caption{Part of the spectrum showing the ratio of a high airmass over a low airmass UVES full-Moon spectra. The oxygen B and A bands at 6900 and 7600\AA\, respectively, are in absorption as expected, but the water vapour bands at 7200 and 8200\AA\ appear in emission, showing that the amount of water vapour along the line of sight was higher when the airmass was lower, indicating that the amount of water vapour above Paranal increased during the first part of the night, before the eclipse. To calculate this ratio, the spectra are aligned by cross-correlation to match the radial velocity of the telluric lines, so the solar lines do not properly cancel out in the ratio, and appear with a wing in emission and the other in absorption. }
    \label{fig:UVES_airmass_raw_ratio}
\end{figure}

We can not proceed exactly as with the HARPS data because full-Moon spectra show that the water-vapour lines were stronger while the Moon was higher in the sky, while it should have been lower for the Moon at a smaller airmass. Fig.~\ref{fig:UVES_airmass_raw_ratio} indeed shows that the amount of water vapour above the telescope increased during the first part of the night, between the two sets of full-Moon observations. We describe below how we consequently corrected our UVES data.

We estimate the precipitable water  vapour (PWV) in our spectra by using the spectral-fitting method described in Querel et al. (2011).  This method compares the measured spectrum with a reference spectrum while varying the reference water-vapour content.  The reference spectrum is generated using an atmospheric model configured with the local environmental parameters at the time of measurement (ambient temperature, pressure, elevation, etc.).  A comparison is then made over several small spectral intervals ($\approx2-5$ nm) where the only significant features are caused by absorption by water vapour.  A least-squares fit is performed between the measured data and reference spectra for a range of PWV values.  The reference spectrum with the smallest fit residual represents the water-vapour content (or PWV) most similar to that of the measured spectrum.  Since this absorption is along the line-of-sight and assuming an isotropic distribution of water vapour above the telescope, the absorption must then be corrected back to zenith by dividing the best-fit PWV by the airmass of the measurement.
 
Tab.~\ref{tab:PWV} indicates that three hours before the eclipse started the PWV measured on the full-Moon spectrum was about 6.4~mm for an airmass of 2. The PWV for airmass=1 was therefore 3.2~mm. 
Later in the night, and just before the penumbra ingress, we find that PWV increased to 8.00~mm at airmass=1.56, indicating the increase of moisture above the telescope. The reduced PWV for airmass=1 is thus $8.00/1.56=5.1$~mm before the eclipse started. Immediately after the eclipse observations, we observed HD80170 and measured a reduced  PWV of 5.71~mm. We therefore assume that during the eclipse, the PWV at zenith remains within the [5.1;5.7]~mm range and we chose the mean value 5.4~mm as the reference PWV at airmass=1 during the eclipse.  PWV values of 5~mm are relatively high for Paranal, and we understand that assuming such a high water vapour content is isotropic and constant above the telescope for three hours (the duration of our observations) may be a strong assumption. Nevertheless, it is the best we can do and the whole process of data reduction leading to $h(\lambda)$ assumes the stability of the PWV above Paranal during the observations (this is obviously also true for the HARPS observations at La Silla).

To correct our eclipse spectra for the signatures of the atmosphere above the telescope, we need distinct reference transmission spectra, one for water and and one for oxygen. In the following, we describe how these reference spectra are obtained. These reference transmission spectra do not need to include Rayleigh and ozone signatures because they are already corrected for by UVES reduction pipeline. 

The UVES transmission water-vapour spectrum $T_{H_2O}(\lambda)$ for airmass=1 (PWV=5.4~mm) is obtained by the ratio of the full-Moon spectra $F$ for which we know the value of PWV (Tab.~\ref{tab:PWV}),
\begin{equation}
  T_{H_2O}(\lambda) =  \bigg( \frac{F(PWV=8.00)}{F(PWV=6.43)} \bigg) ^{5.4/(8.00-6.43)}. \label{ratio_H2O}
\end{equation}

To derive a spectrum for pure water vapour (\textit{i.e.} without the lines from other elements or molecules), $T_{H_2O}(\lambda)$ is calculated only when water-vapour absorption is present ($T \leq 99\%$) in the water-vapour synthetic spectrum computed to measure the PWV. To avoid $T_{H_2O}(\lambda)$ to become too noisy in the deepest  absorption bands of water vapour, the spectrum is not calculated when water-vapour transmission becomes $T \leq 2.5\%$.

\begin{table}
  \caption{Measured PWV before the eclipse and at the end the observation, and reduced PWV for airmass=1.}
  \label{tab:PWV}
  \begin{center}
    \leavevmode
    \begin{tabular}{lllll} \hline \hline              
  h:min UT              & PWV     & Airmass   &   PWV  (mm)                      &  Note \\  
                                &       (mm)                 &                  &  at zenith                      &            \\  \hline 
  02:05 and 02:18 & 6.43              & 2.01           &     	3.20				& 1 \\
  05:08 and 05:21 & 8.00              & 1.56           &     	5.14				 & 2\\
  08:23 and 08:32 & 5.95              & 1.04          &     	5.71				 & 3\\  \hline
  \multicolumn{5}{l}{}                                             \\       
  \multicolumn{5}{l}{1: full-Moon before penumbra ingress, DIC2 and DIC1} \\ 
  \multicolumn{5}{l}{observations, respectively.}         \\
  \multicolumn{5}{l}{2: full-Moon before penumbra ingress, DIC2 and DIC1} \\ 
  \multicolumn{5}{l}{observations, respectively. First contact with penumbra } \\ 
  \multicolumn{5}{l}{ occurred at about 05:28 UTC.}         \\
  \multicolumn{5}{l}{3: HD80170 at the end of the eclipse observation, DIC1 } \\ 
  \multicolumn{5}{l}{and  DIC2 observations, respectively.} \\

    \end{tabular}
  \end{center}
\end{table}

The transmission spectrum for $O_2$ for airmass=1 is obtained by taking into account the variation of airmass AM between the two $F$ spectra (Tab.~\ref{tab:PWV}), as done by  Vidal-Madjar et al. (2010):
\begin{equation}
  T_{O_2}(\lambda) =  \bigg( \frac{F(AM=2.01)}{F(AM=1.56)} \bigg) ^{1/(2.01-1.56)}. \label{ratio_O2}
\end{equation}
And as for $T_{H_2O}$,  $T_{O_2}$ is calculated only when $O_2$ absorption is present ($99.9\% \geq T\geq 2\%$) in a $AM=1$ MODTRAN\footnote{http://modtran5.com/} synthetic spectrum.

\begin{figure}[h]
   \centering
   \includegraphics[width=9cm]{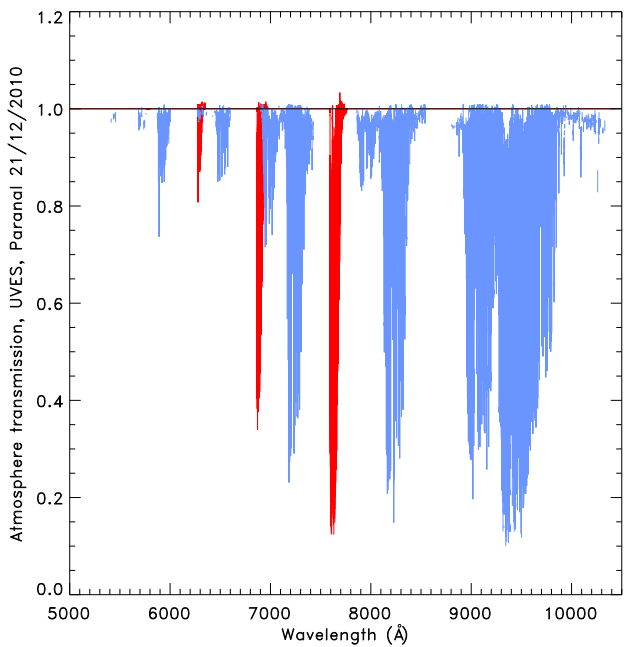}
   \caption{Paranal atmosphere transmission function $T(\lambda)$ for AM=1 as evaluated
   from full-Moon spectra taken with UVES before the eclipse, following the reduction process described in Sect.~\ref{section_UVES_trans}. The UVES reduction pipeline automatically removes from all individual spectra tabulated values for the Rayleigh, aerosol and ozone signatures at the corresponding airmass of the observation. These signatures therefore do not appear here as for HARPS (Fig.~\ref{transmission}). A weak residual curvature of the continuum ($\approx5\%$ maximum at 7800\AA, with no resemblance to a residual signature of ozone or Rayleigh) was fitted with a second-order function. The spectrum shown here is normalized to this fit to have a flat continuum at 1 (horizontal line).  The O$_2$ $\gamma$, A, and B bands are clearly seen (in red), the other bands are due to H$_2$O (in blue). }
    \label{fig:UVES_trans}
\end{figure}

Once $T_{H_2O}$ and  $T_{O_2}$ are known, we can correct our spectra simply by the division
\begin{equation}
  s =  \frac{S}{(T_{H_2O}\times T_{O_2})^{AM}} \label{corr_AM},
\end{equation}
where $s$ is the corrected spectrum and $S$ is a full-Moon or eclipse spectrum acquired at airmass $AM$. The final $T(\lambda)$ spectrum for Paranal is shown in Fig.~\ref{fig:UVES_trans}. The continuum shows neither the Rayleigh nor the ozone signature in this spectrum (on the opposite of what is obtained with HARPS, Fig.~\ref{transmission}) because the UVES reduction pipeline automatically removed from all spectra tabulated values for the Rayleigh and ozone signatures at the corresponding airmass of the observation. Of course, the additional Rayleigh and ozone signatures present in the eclipse spectra and produced by the light path through the limb of the Earth atmosphere have not been removed by the UVES pipeline.

To calculate $T_{H_2O}$ and  $T_{O_2}$, the full-Moon spectra are aligned with a cross-correlation to match the radial velocity of the telluric lines.

\section{Results and discussion}
\label{Results}
\subsection{Results from Eq.~\ref{ratioAB} (method1)}
From the HARPS data, we build five pairs of penumbra spectra suitable for Eq.~\ref{ratioAB}, [98;204], [125;232], [66;125], [66;204], and [66;232] where the numbers in brackets are the positions described inTab.~\ref{tab:HARPS_fiber_pos}). The possible pair [66;98] is not used: it produces a significantly noisier altitude profile probably because the positions in the penumbra are too close and the difference between the spectra is too small.

The penumbra spectra are affected by the wavelength-dependent solar limb darkening $LD(\lambda)$ (\cite{hestroffer1998}).  Eq.\ref{ratioAB} makes use of geometrical parameters $S$ and $L$ for two different solar crescents visible above Earth. The solar $LD(\lambda)$ makes the crescents fainter than the mean solar brightness. The size of the solar crescents estimated from the penumbra spectra fluxes is thus underestimated if the penumbra fluxes are not corrected for the solar $LD$.  
To correct for this effect, we first evaluate the size of the solar crescents (parameters $S$ and $L$) from the flux ratio (penumbra/full-Moon) at  $\lambda$=5798.8\AA\ which allows us to evaluate an average $LD$ that affects each solar crescent following Hestroffer \& Magnan (1998).
For these crescent sizes and the $LD$ obtained at $\lambda$=5798.8\AA, a wavelength-dependant limb darkening $LD(\lambda)$ is then re-evaluated from the low spectral resolution model of Hestroffer \& Magnan (1998). It covers the spectral ranges [3033-3570] and [4160-10990]\AA, with a spectral resolution of a few tens (we do a linear interpolation between 3570 and 4160\AA). The model is then used to calculate the $LD$-corrected penumbra spectra $E_A(\lambda)/LD(\lambda)$ and $E_B (\lambda)/LD(\lambda)$ from which new $S$ and $L$ are evaluated for $\lambda_0$ with Eq.~\ref{S}. Eq.\ref{ratioAB} is then evaluated again to derive the final $h(\lambda)$ profile.

\begin{figure}
   \centering
   \includegraphics[width=9cm]{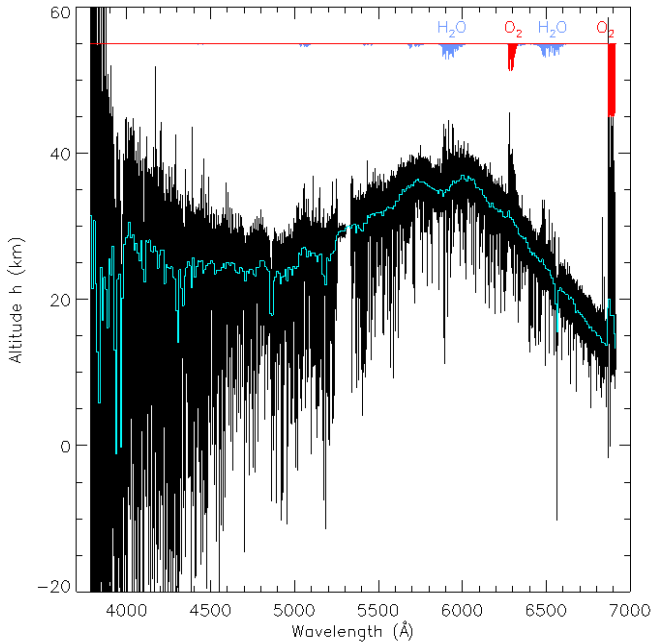}
   \caption{Altitude profile with HARPS as obtained from Eq.~\ref{ratioAB} (method~1).  A 10\AA\ bin is superimposed in light-blue over the full-resolution profile. The profile shows the
   large bump due to ozone Chappuis band around 6000\AA, the oxygen $\gamma$- and B-band at 6300 at 6900\AA\, respectively, and also the water-vapour band around 6500\AA. The expected increased of $h(\lambda)$ in  the blue due to Rayleigh is visible as a negative slope between 4000 and 5000\AA. Oxygen and water-vapour absorption spectra at arbitrary scale are shown.}
    \label{fig:h_mean_HARPS_noRingcorr}
\end{figure}

\begin{figure}
   \centering
   \includegraphics[width=9cm]{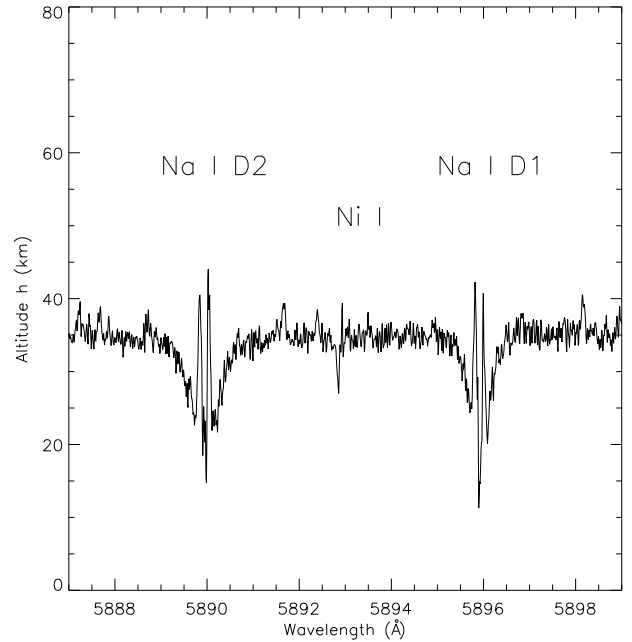}
   \caption{Zoom in the Na I doublet region of the altitude profile  from Fig.~\ref{fig:h_mean_HARPS_noRingcorr} with HARPS and method~1. The $\approx 2$ \AA -wide photospheric wings of the Na lines are still present,    showing that the solar signature is not properly removed from this $h(\lambda)$ profile. A residual solar line of Ni I is visible at $\approx 5892.9$\AA\ (\cite{wallace2011}).}
    \label{fig:h_mean_HARPS_noRingcorr_zoomNa}
\end{figure}

The mean altitude profile from the five pairs of fibers obtained from Eq.\ref{ratioAB} is shown in Fig.~\ref{fig:h_mean_HARPS_noRingcorr}. The profile is shifted upward by $\approx30$~km to have an altitude of 23.8~km in the [4520-4540]\AA\ domain, as predicted by Ehrenreich et al. (2006). The profile shows the
large bump due to ozone Chappuis band centered around 5900\AA\ and peaking here at about 38~km, and the oxygen $\gamma$ and B bands at 6300 and 6900\AA, respectively. Water vapour also appears around 6500\AA\  at an altitude above 25~km. The increase of $h(\lambda)$ in the blue due to Rayleigh scattering is visible as a negative slope between 4000 and 5000\AA (\cite{ehrenreich2006}; \cite{kaltenegger2009}).  The blue part of the profile shows many residual solar absorption lines that bias the binned profile towards lower altitudes. Numerous solar lines are visible, for example H$\beta$ at 4861\AA\ and H$\alpha$ at 6563\AA.
The Fig.~\ref{fig:h_mean_HARPS_noRingcorr_zoomNa} is a zoom on the Na I doublet. The $\approx$~2\AA-wide photospheric wings of the solar Na lines are still visible. The presence of all these solar features indicates that the solar signature is not properly removed from the $h(\lambda)$ profile. 
This could result from the $E/F$ ratios in Eq.\ref{ratioAB}. The penumbra spectra $E$ indeed  contain the spectrum from the solar crescent, that is, the edge of the Sun where solar absorption lines are deeper than in full-Moon spectra $F$, which correspond to a full disk of the Sun illuminating the Moon. We therefore tried to adjust a power $p>1$ on $F$ to compensate for the solar residuals with ratios $E/(F^p)$ in Eq.\ref{ratioAB} instead of   $E/F$. But we found $p$ to be very close to 1 and could not detect a significant difference between the average depth of the absorption lines in $E$ and $F$.

Vidal-Madjar et al. (2010) observe the same type of solar residuals in the SOPHIE data (residual wings of the Na I doublet) and the explanation invoked is the Ring effect. The depth of solar Fraunhofer lines in scattered light is lower than in direct sunlight. This effect is known as the Ring effect (\cite{grainer1962}) and is mainly induced by rotational Raman scattering (\cite{vountas1998}). The line filling-in for forward scattering (as experienced by the photons that interact with the Earth atmosphere and are scattered towards the eclipsed Moon) is about 2.1\% and 3.8\% of the continuum for single and multiple scattering, respectively, for the Ca II K line (\cite{vountas1998}). Kattawar et al. (1981) report a filling-in of 2.5\% at 6301.5\AA\ for a Fe-I line. Brinkmann (1968) calculates a filling-in of about 1.7 in the center to 1.9\% in the wings for a hypothetical line at 4000\AA, underlining the continuous and smooth effect across the line profile, which is also measured by Pallamraju et al. (2000). Brinkmann (1968)  reports observations showing a decrease of the filling-in with wavelength by a factor of 2 between 4383 and 6563\AA, and more recent measurements (\cite{pallamraju2000}, Fig. 3 therein) shows that the Ring effect decreases approximately linearly by a factor of $\approx3$ from 4285 to 6560\AA\  for a given line strength (normalized depth to continuum $\times$ width). 
 
Several methods exist to correct for the Ring effect (\cite{wagner2001}). The simplest is that of Noxon et al. (1979), which  consists of subtracting a small constant, close to 2\%, from the spectrum.  This was the solution applied in  Vidal-Madjar et al. (2010) where a constant is empirically adjusted to remove the residual solar Na I lines (the constant was 4\% and 0.6\% of the continuum in the Na I region, for the deepest and the shallowest positions in the penumbra, respectively). We also correct this Ring effect in our data.

We observed that removing a Ring value $x_a$ from the shallowest spectrum in the penumbra and $x_b (>x_a)$ to the deepest spectrum gives almost the same result as removing $x_b-x_a$ to the deepest only (all spectra were corrected from the instrumental response, so Noxon's approach applies correctly). The Ring signature is thus, as expected, stronger in the deeper spectrum ($x_b > x_a$), where collected photons  have been more scattered in Earth atmosphere than photons collected in the shallower penumbra. 
In our empirical correction for the Ring effect, we therefore only apply a correction on the spectrum in the deepest position in the penumbra. We consider empirical  Ring corrections following 
\textit{i)} Noxon et al. (1979), 
\textit{ii)} Noxon et al. (1979) except in deep telluric bands (T $<$ 0.5) where the Ring effect is negligible because Raman-scattered photons are strongly absorbed (\cite{sioris2000}), 
\textit{iii)} a  $\lambda^{-2}$ power law to mimic the chromatic dependance observed by  Pallamraju et al. (2000) or Brinkmann (1968),
\textit{iv)} a  $\lambda^{-2}$ power law subtracted except in deep telluric bands (T $<$ 0.5), and 
\textit{v)} a subtraction of the reciprocal of the spectrum (\cite{jmck1989}) that mimics the Ring spectrum. Here again, the obtained Ring spectrum is not removed from deep telluric lines (T $<$ 0.5).  

\begin{figure}
   \centering
   \includegraphics[width=9cm]{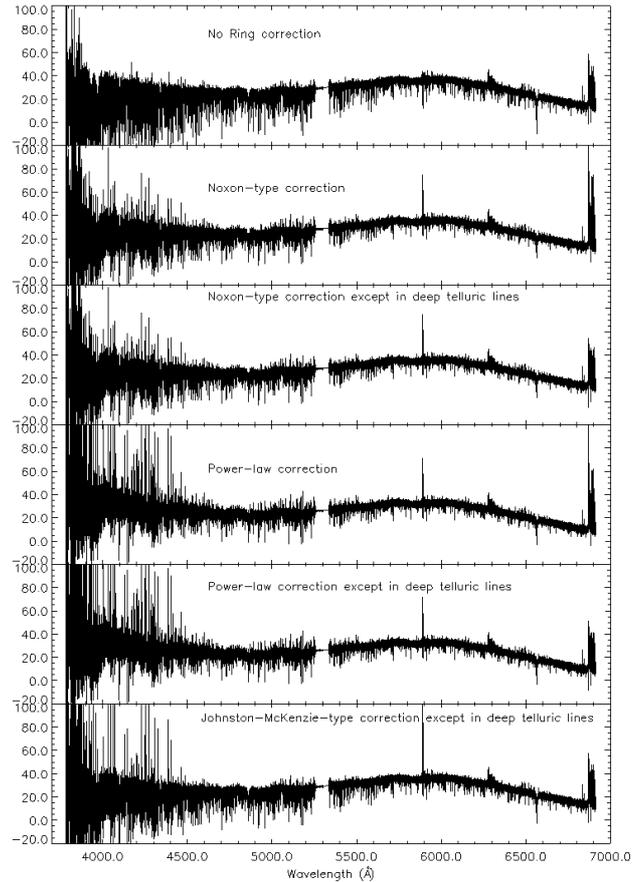}
   \caption{Altitude profile with HARPS and method~1 without Ring correction (upper panel) and different empirical Ring corrections. The amount of correction is $1\% $ of the continuum at 5880\AA: \textit{i)} Noxon et al. (1979), 
\textit{ii)} Noxon et al. (1979) except in deep telluric bands (T $<$ 0.5) where the Ring effect is negligible because Raman-scattered photons are strongly absorbed (\cite{sioris2000}), 
\textit{iii)} a  $\lambda^{-2}$ power law to mimic the chromatic dependance observed by  Pallamraju et al. (2000) or Brinkmann (1968),
\textit{iv)} a  $\lambda^{-2}$ power law subtracted except in deep telluric bands (T $<$ 0.5), and 
\textit{v)} a subtraction of the reciprocal of the spectrum (\cite{jmck1989}), which mimics the Ring spectrum. Here again, the obtained Ring spectrum is not removed from deep telluric lines (T $<$ 0.5).  
The Na~I doublet peaks at $\approx$~5890\AA\ for all types of Ring corrections tested.}
    \label{ring_all}
\end{figure}

\begin{figure}
   \centering
   \includegraphics[width=9cm]{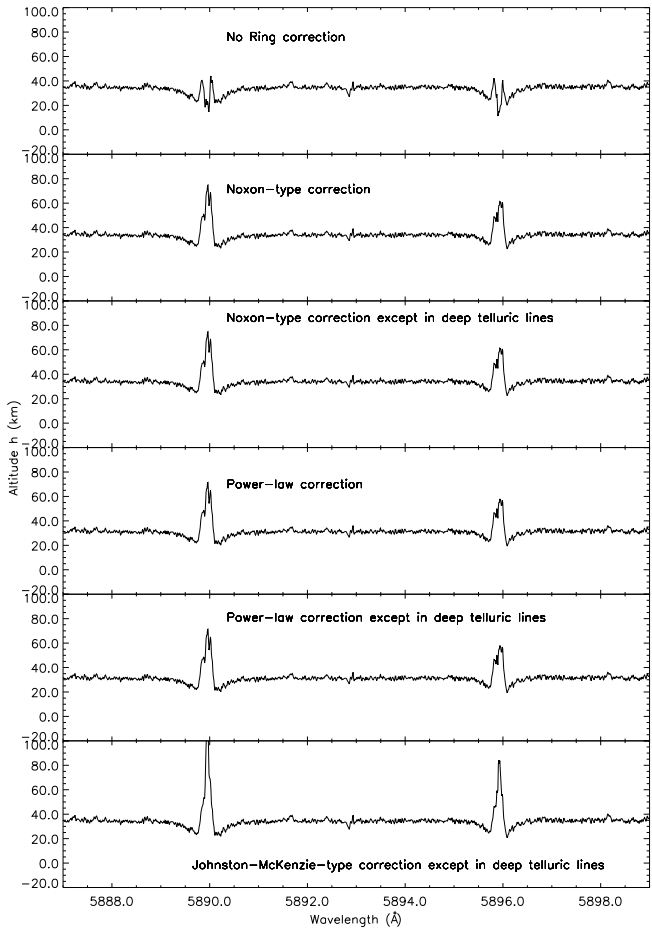}
   \caption{Same as  Fig.~\ref{ring_all} but zoomed-in on the Na I doublet region of the altitude profile without Ring correction (upper panel) and different empirical Ring corrections.  }
    \label{ring_all_zoomNa}
\end{figure}

We observe that removing $\approx 1\% $ of the continuum at 5880\AA\ to the deepest spectrum of each pair reduces the solar residuals over the full altitude profile (Fig.~\ref{ring_all}) and slightly reduces the Na wings (Fig.~\ref{ring_all_zoomNa}). The correction also leads to the rise of the Na lines to altitude values of $\approx 80$~km, a value in agreement with the literature (\cite{plane2003}; \cite{moussaoui2010}). Nevertheless, the width of the terrestrial Na I peaks  are observed to be $\approx0.4$\AA. This is larger than the $\approx0.2$\AA\ observed for the other telluric lines, but might be induced by the presence of the strong Na I line in the solar spectrum. The extraction process of $h(\lambda)$ is then more disturbed in the case of the Na I lines, easily inducing a change in the width, particularly because the solar lines cannot be perfectly positioned in all spectra since these have been Doppler-shifted to have the telluric lines aligned. An additional effect is the limb-darkening effect that changes the shape of the solar Na I lines emitted by different size crescent above the Earth limb. This is also why the broad Na I stellar wings are not perfectly corrected in the presented extraction procedure, which only includes a low spectral resolution limb-darkening model.   Finally, we also note that only the Na I lines are detected in Fig.~\ref{ring_all}, not the strong Mg I triplet at $\approx$~5170\AA\ nor H$\alpha$, for example, produce a peak in the altitude profile. We therefore conclude that the terrestrial Na layer is detected. 

Interestingly, Garc\'ia-Mu\~noz et al. (2012) point out that the lunar eclipse of 16 August 2008 occurred only four days after the maximum of the Perseid meteor shower. It is indeed known that meteor showers can temporarily increase the amount of sodium in the atmosphere by a factor up of 10 (\cite{plane2003}; \cite{moussaoui2010}). This corroborates the detection of the sodium layer by Vidal-Madjar et al. (2010) in August 2008. Nevertheless, the Geminid meteor shower maximum occurs on December 13. It is an active meteor shower, with a particle density similar to that of the Perseids\footnote{See for example the IMO (International Meteor Organization) data at http://www.imo.net/live/geminids2012/ and http://www.imo.net/live/perseids2012/.}, that might have increased the sodium absorption during the eclipse of 21 December 2010. Nevertheless, our observations just show that the Na layer is indeed present, as during our first observations in August 2008. Since we have only one measurement, it is unfortunately  not possible to detect any change of the atmospheric Na I content, either in 2008 or in 2010, and thus to confirm a link with any meteor shower.

\begin{figure*}[h]
   \centering
   \includegraphics[width=18cm]{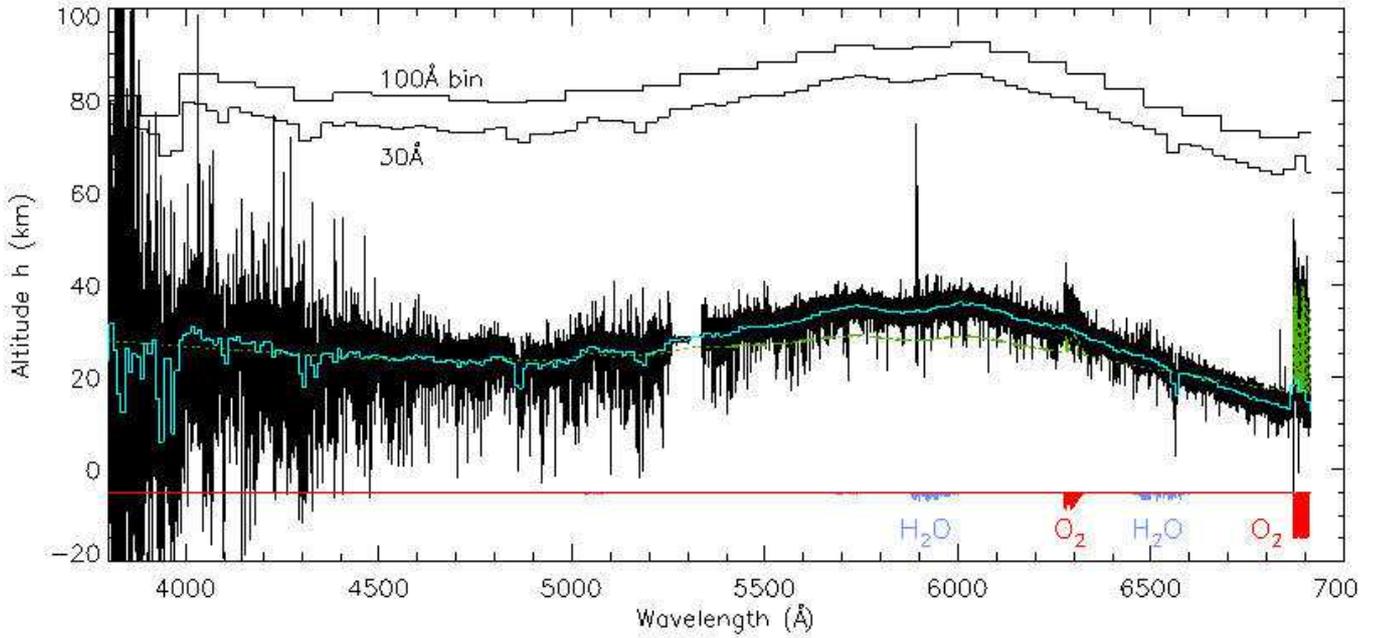}
   \caption{Final altitude profile obtained with HARPS, method~1, and with a Noxon correction for the Ring effect, expect in the deep telluric lines (\cite{sioris2000}). A 10\AA\ bin is superimposed in light-blue over the full-resolution profile; 30 and 100\AA\ bins are shown too, shifted upwards for clarity. The Na~I peaks are visible near 5900\AA.  Oxygen and water-vapour absorption spectra at arbitrary scale are shown. The altitude profile has been shifted to have an altitude of 23.8~km in the [4520-4540]\AA\ domain, as predicted by Ehrenreich et al. (2006) (dots and dashes in green - dashes to improve the visibility of the curve around 6000\AA). }
    \label{HARPS_full_final}
\end{figure*}

\begin{figure}[h]
   \centering
   \includegraphics[width=9cm]{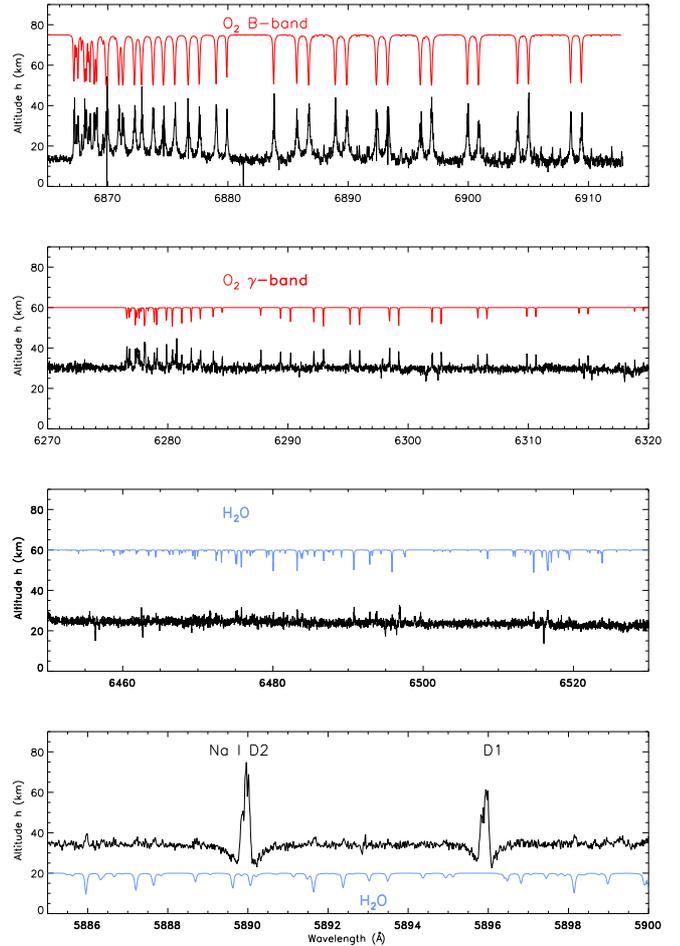}
   \caption{Zooms on the HARPS final altitude profile from method~1 (Fig.~\ref{HARPS_full_final}) with a Noxon correction for the Ring effect, except in the deep telluric lines (\cite{sioris2000}). Oxygen and water vapour absorption spectra at arbitrary scale are shown. Oxygen shows up very clearly. Water vapour is barely visible around 6500\AA\ above 25~km and  weak lines are visible in the higher Na I D1 D2 region. Note that our Noxon correction does not completely remove the wings of the Na~I solar lines. The wings disappear with a stronger correction, but the two Na~I peaks then reach an altitude well above 100~km which is not realistic.}
    \label{HARPS_full_final_zoom}
\end{figure}

It can be seen from Fig.~\ref{ring_all} that the Ring correction is not homogeneous across the full spectral range: The $H\alpha$ line remains undercorrected, while lines in the bluer part of the $h(\lambda)$ profile are overcorrected. We therefore adopt the simplest Noxon approach, a subtraction of a constant, except in deep telluric lines (T $<$ 0.5). The final result and the model from Ehrenreich et al. (2006) are shown in Fig.~\ref{HARPS_full_final}. The model is flatter than the observed profile, which peaks at about 35~km near 6000\AA, about 8~km above the model. This difference might be a refraction effect that amplifies the observed profile, as with method~3, although this effect is weaker here because we use only penumbra spectra. If the profile is calculated only with the two pairs [98;204] and [125;232] (Tab.~\ref{tab:HARPS_fiber_pos}),  leaving out the spectrum from the deepest penumbra at only 66 arcsec from the umbra, the observed profile is indeed flatter (but noisier): it still peaks at 35~km, higher that the altitude expected from the model, but the lowest altitude at 6900\AA\  becomes $\approx17$~km instead of 12.   We note that our observations of the 2008 eclipse agrees better  with the model (\cite{vidal-madjar2010}, Fig.9 therein) although SOPHIE fibers were between $\approx$~100 and 250~arcsec from the umbra edge, as for HARPS.
The additional bins  in Fig.~\ref{HARPS_full_final} show that the B-band is barely detectable in the 30\AA\ bin, while the broad O$_3$ Chappuis band and Rayleigh slope in the blue remain visible in the 100\AA\ bin.
Fig.~\ref{HARPS_full_final_zoom}  shows zooms in the Na I lines and the O$_2\ \gamma$ and B band regions. Oxygen shows up very clearly, with peaks reaching h=35-40~km. Some water-vapour peaks are visible but weak, which is not surprising because of the high altitude at which  they are detected and also because of the lower strength of the water-vapour absorption in the HARPS wavelength range compared with $\lambda> 7000$\AA.  We detect  neither Li (6708\AA) nor Ca II  (\cite{hunten1967}; \cite{plane2003}).

\begin{figure}
   \centering
   \includegraphics[width=9cm]{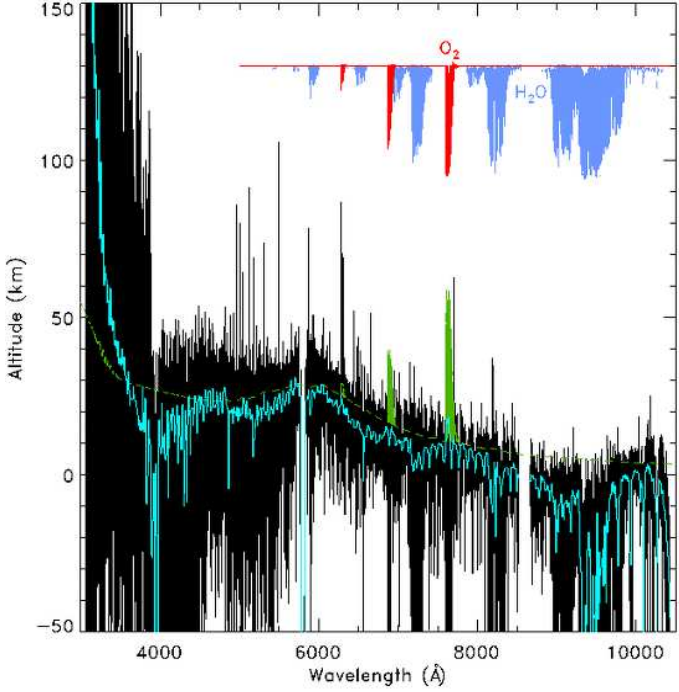}
   \caption{Final altitude profile obtained with UVES and method~1. No Ring correction is applied. A 10\AA\ bin is superimposed in light-blue on the full-resolution profile. Reference oxygen and water-vapour absorption  spectra obtained with UVES are shown at arbitrary scale. The altitude profile has been shifted to have an altitude of 23.8~km in the [4520-4540]\AA\ domain, as predicted by Ehrenreich et al. (2006) (green dashed line, no water vapour in the model). }
    \label{UVES_methodSOPHIE}
\end{figure}

With UVES, the coordinates of the telescope indicate that the telescope unfortunately did not move in the penumbra between the exposures in order to record spectra at 90 and 180 arcsec from the umbra edge as expected, but instead stayed $\approx200$  arcsec from the umbra  (Tab.~\ref{tab:UVES_slit_pos}). The different recorded spectra are very similar and the result from Eq.~\ref{ratioAB} is noisy, especially in the blue. Residuals from spectral orders are also visible above  7000\AA. There is nevertheless a spectrum taken in UVES DIC2 configuration, while the telescope was thought to be in the umbra, but was too close to the Moon edge and left the moon about 20 s after the exposure started. The exposure was then stopped, but the telescope coordinates indicate that the telescope was in fact in the penumbra. The exposure time for that unexpected exposure is only about 21 and 38\% of what was required for the BLUE and RED exposures, respectively.  The altitude profile obtained with UVES is shown in Fig.~\ref{UVES_methodSOPHIE}. The overall shape agrees with the model of Ehrenreich et al. (2006), but the altitude observed in the UV is far higher than that of the model. Moreover, we had to shift the DIC1 BLUE profile by -110~km to connect it to the DIC2 BLUE profile. The DIC2 RED profile  has also to be shifted by 35~km to be connected to the altitude obtained for the DIC1 RED profile. The oxygen bands are barely visible and the water-vapour bands appears in absorption. We did not consider any Ring correction for this noisy profile. Note also that the model of Ehrenreich et al. (2006) shown in the figures of this paper does not include water vapour. 

\subsection{Results from Eq.~\ref{ratioA} (method~2)}
\label{result-method2}

\begin{figure}[h]
   \centering
   \includegraphics[width=9cm]{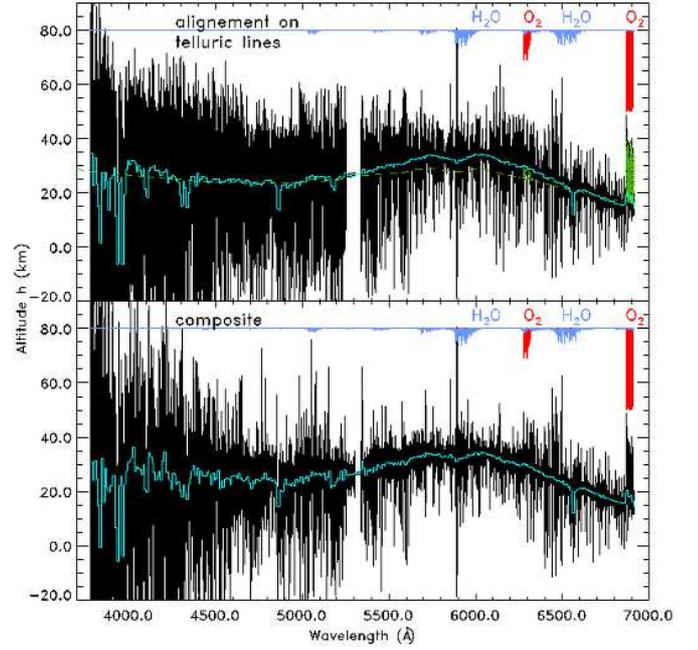}
   \caption{Altitude profile with HARPS and Eq.~\ref{ratioA} (method~2). Only the spectrum of the deepest penumbra, taken at 66~arcsec from the umbra, is used here (Tab.~\ref{tab:HARPS_fiber_pos}). Shallower penumbra spectra give similar but noisier profiles. A 10\AA\ bin is superimposed in light-blue on the full-resolution profiles. The upper panel shows the profile obtained with the HARPS spectra aligned on the telluric lines. Solar lines are consequently slightly misaligned and generate noise in the profile. The lower panel shows a composite profile showing the profile calculated with the spectra aligned on the telluric lines, expect when a solar line is present, where the profile from spectra aligned on the solar lines is used. Oxygen and water-vapour absorption spectra at arbitrary scale are shown. The green dashed line is the model of Ehrenreich et al. (2006).}
    \label{HARPS_Eq3_1spectre}
\end{figure}

\begin{figure}[h]
   \centering
   \includegraphics[width=9cm]{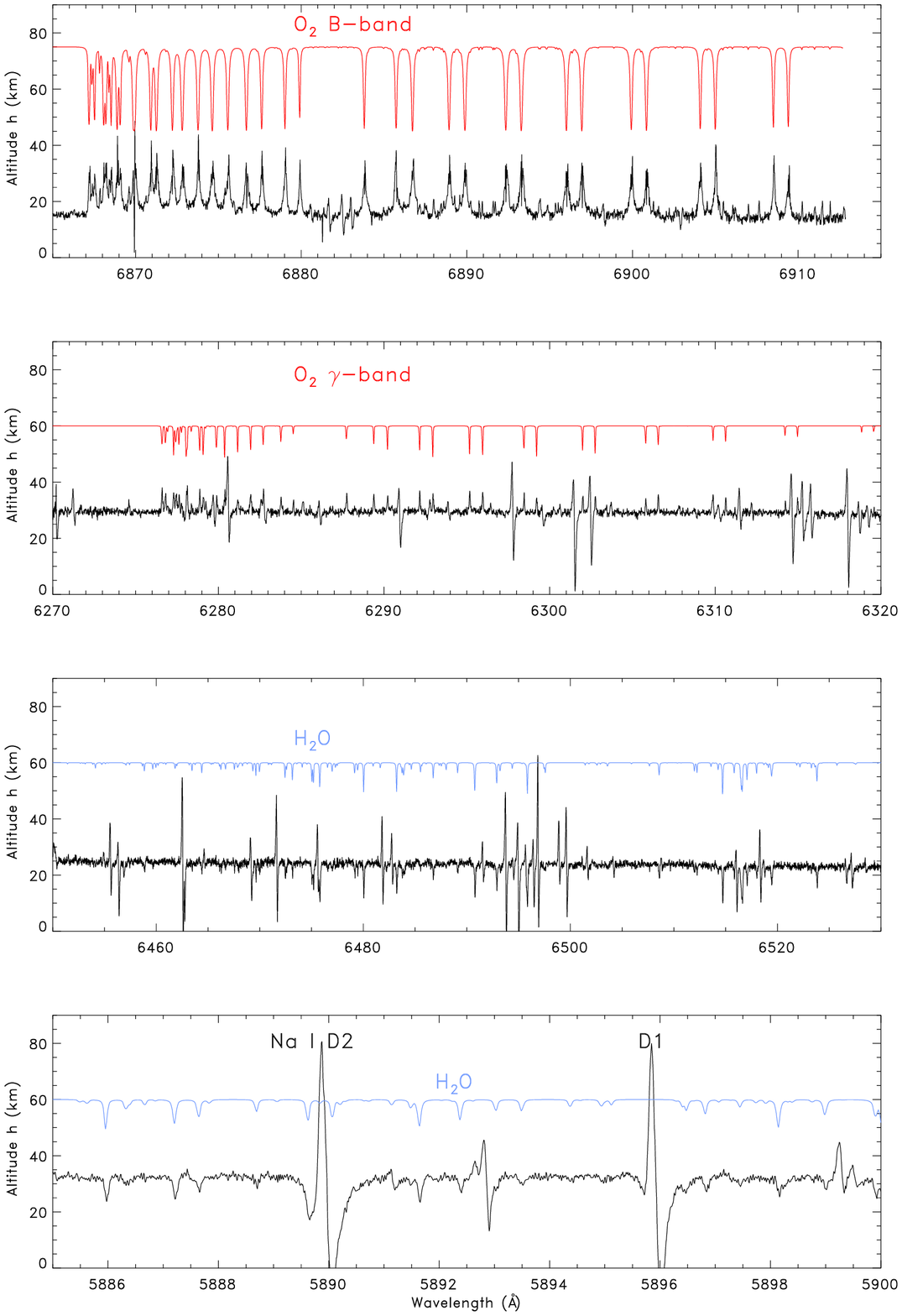}
   \caption{Zooms on the HARPS altitude profile from method~2 shown in Fig.~\ref{HARPS_Eq3_1spectre}. Oxygen and water-vapour absorption spectra at arbitrary scale are shown to facilitate the identification of the spectral lines. Oxygen is correctly detected, while water vapour is not detected. Atmospheric Na I D-lines are not detected either, maybe they are hidden by the strong solar residual lines.}
    \label{HARPS_Eq3_1spectre_zoom}
\end{figure}

The HARPS spectra are now used with Eq.~\ref{ratioA} to calculate $h$ from only one penumbra spectrum. The eclipse spectra are corrected  for limb-darkening and Ring effect as for method~1. When the profile is calculated with the spectra in Eq.~\ref{ratioA} aligned on the telluric lines, the solar lines become consequently slightly misaligned, which generates noise in the profile (Fig.~\ref{HARPS_Eq3_1spectre} upper panel). We therefore build a composite profile where all lines - telluric and solar - are aligned. The composite uses two different profiles, one aligned on the telluric lines and a second aligned on the solar line. In the first profile, every misaligned solar line is recognized with the help of a synthetic G2 spectrum and replaced by its aligned residual taken from the second profile.  Some solar lines are missing in the G2 spectrum in the most crowded telluric bands, therefore some residual noise due to uncorrected solar lines  is still present in the composite inside these telluric bands. The procedure still slightly cleans the $h$ profile (Fig.~\ref{HARPS_Eq3_1spectre} lower panel). The overall shape of the profile is similar to the profile obtained with Eq.~\ref{ratioAB}  shown in Fig.~\ref{HARPS_full_final}. Ozone appears about 6~km above the model. When zooming-in (Fig.~\ref{HARPS_Eq3_1spectre_zoom}), only the oxygen B and $\gamma$ bands are correctly detected with peaks reaching h=35-40~km. Water-vapour bands at 6500 and 5890\AA\  are not visible, nor the atmospheric Na I D-lines, which are maybe hidden in the strong solar residual lines. Note that the profile shown here is for the deeper position in the penumbra. The other positions lead to a noisier result, although the overall shape is correct.

\begin{figure}[h]
   \centering
   \includegraphics[width=9cm]{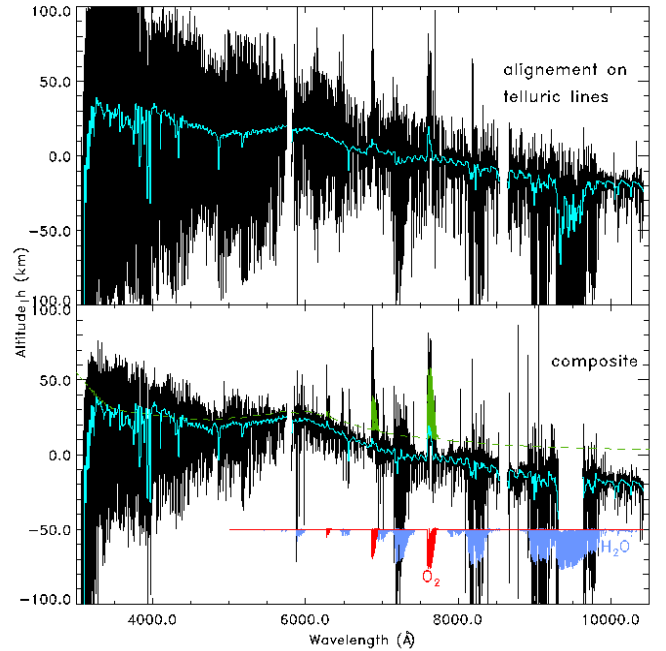}
   \caption{Altitude profile with UVES and method~2. A 10\AA\ bin is superimposed in light-blue on the full-resolution profile. The upper panel shows the profile obtained with the UVES spectra aligned on the telluric lines, with solar lines consequently slightly misaligned and thus generating noise in $h$. The lower panel shows a composite profile showing the profile aligned on the solar lines when no telluric line is present, and the profile aligned on the telluric lines when a telluric line is present. Oxygen and water-vapour absorption spectra at arbitrary scale are shown to facilitate the identification of the spectral lines. The model of Ehrenreich et al. (2006) is shown in green.  }
    \label{UVES_Eq3_1spectre}
\end{figure}

\begin{figure}[h]
   \centering
   \includegraphics[width=9cm]{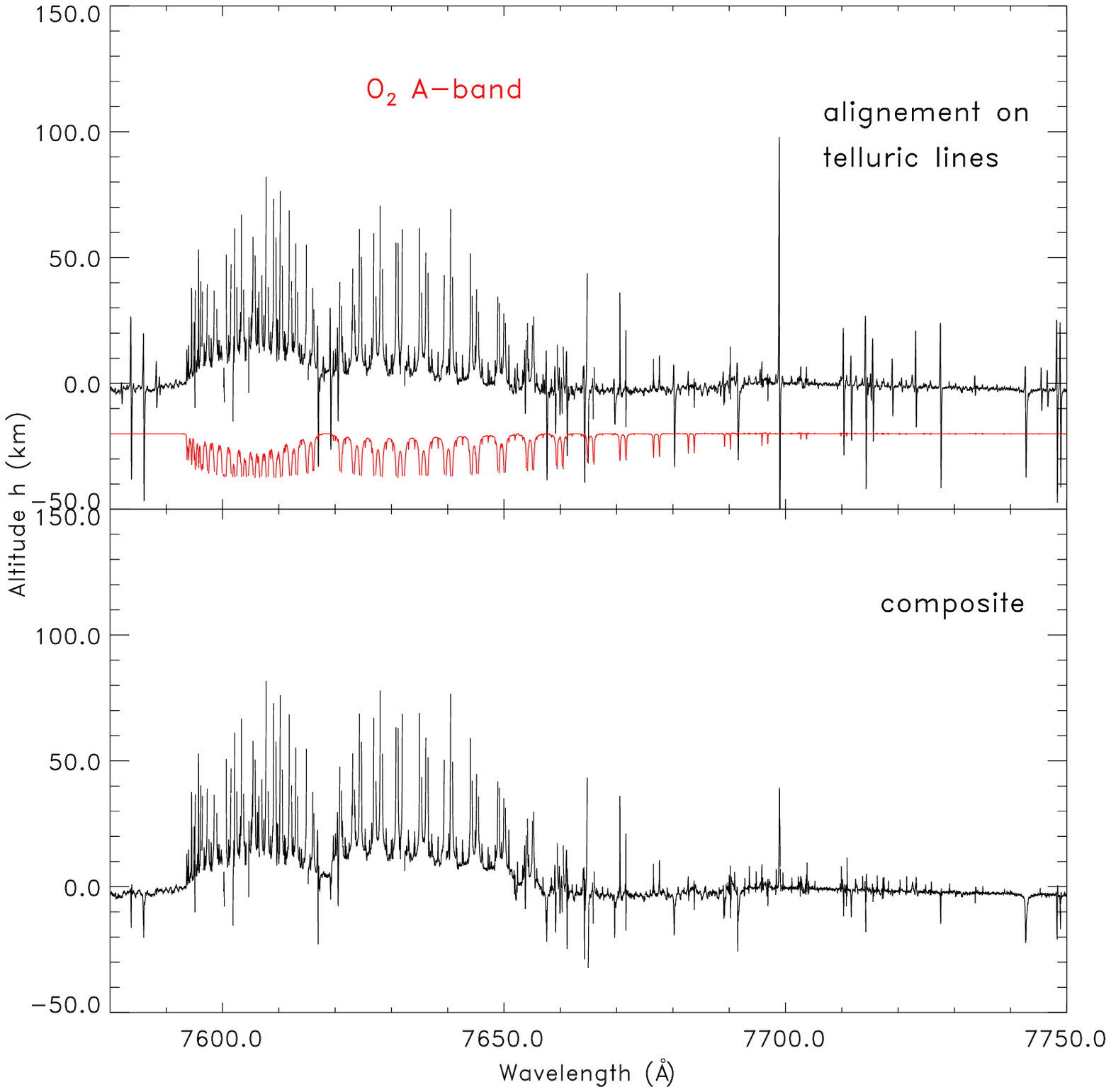}
    \includegraphics[width=9cm]{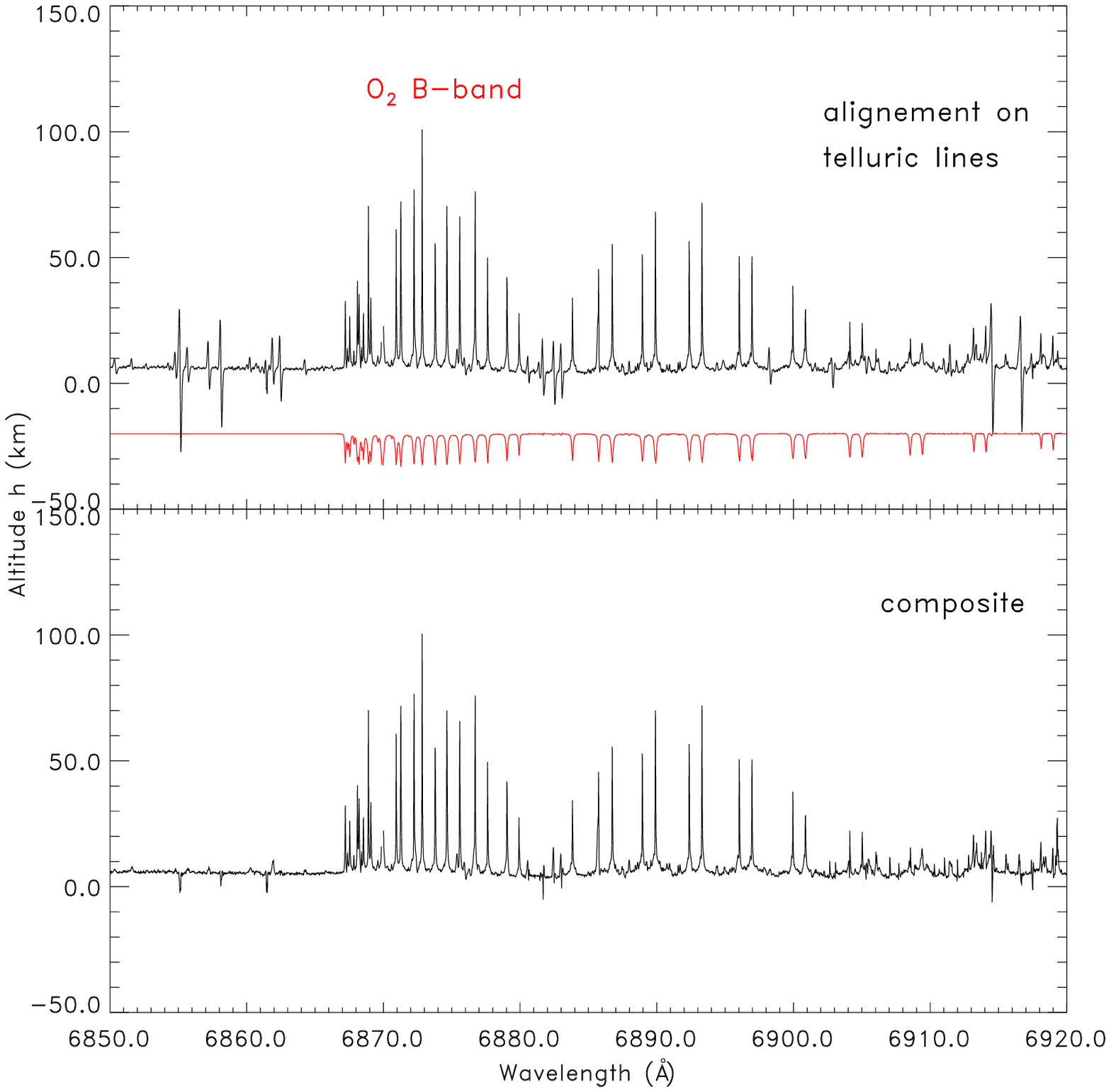}
   \caption{Zooms on the oxygen A and B-band regions of the UVES profil shown in Fig.~\ref{UVES_Eq3_1spectre} (method~2). An oxygen absorption spectrum at arbitrary scale is shown.}
    \label{UVES_Eq3_1spectre_zoomO2}
\end{figure}

Eq.~\ref{ratioA}  is also applied on UVES spectra to calculate $h$. As for HARPS, we compute a composite profile. The composite is the profile obtained with the spectra aligned on the solar lines in which the oxygen and water-vapour lines are replaced by the oxygen and water-vapour lines picked up in the profile  obtained with the spectra aligned on the telluric lines. This procedure cleans the profile (Fig.~\ref{UVES_Eq3_1spectre} lower panel), although compromising the detection of atmospheric species different from oxygen and water. This may explain, at least partially, why the Na I D-lines do not  appear clearly. The oxygen $\gamma$, B, and A bands are visible up to $\approx80$~km (zooms in Fig.~\ref{UVES_Eq3_1spectre_zoomO2}), but water-vapour bands are not detected. We subtract a Ring effect from the UVES eclipse spectra of 0.5\% of the continuum at 5880\AA. When we subtract a Ring effect of 1\% as we did for HARPS, the profile does not change, except from the oxygen A and B bands which  peak at 150~km.

Qualitatively speaking, the overall shape of the profile from 3000 to 7000\AA\  is in acceptable agreement with the model, with the ozone Chappuis bump at 6000\AA\  and a Rayleigh increase of $h$ in the blue and near UV. The DIC2 BLUE altitude is set to $h$=23.8~km in the [4520-4540]\AA\ domain, as predicted by Ehrenreich et al. (2006). The other parts of the profile need to be shifted by about $\pm6$~km to be properly continuously connected to the DIC2 BLUE profile.


\subsection{Results from the umbra edge radiance profile (method~3) }
\label{sect:result_method3}
With method~3, we make use of all available umbra and penumbra spectra. The penumbra spectra are corrected from the limb-darkening in the same way as for method~1 and 2.  For each sampled wavelength in the HARPS spectra, we fit a sigmoid curve (Eq.~\ref{eq:model}) through the eight available measured points (position, magnitude) in the umbra and deep penumbra region (Tab.~\ref{tab:HARPS_fiber_pos}). An example of the obtained fit is shown in Fig.~\ref{fig:umbrapenumbraHARPS}. We now need to define the radius of the umbra edge $r_e$ from which we build the Earth atmosphere thickness $h(\lambda)$ simply given by
\begin{equation}
 h(\lambda) = r_e(\lambda) \times \frac{r_\oplus }    {r_u}  + cte,
  \label{eq:measured_h}
\end{equation}
where $r_e$ is expressed in arcmin, $r_u=42.7$~arcmin is the umbra radius from the ephemeris, and $r_\oplus=6371$~km is the Earth radius. The constant $cte$ is set to give $h=23.8$~km at [4520-4540]\AA.

\begin{figure}[h]
   \centering
   \includegraphics[width=9cm]{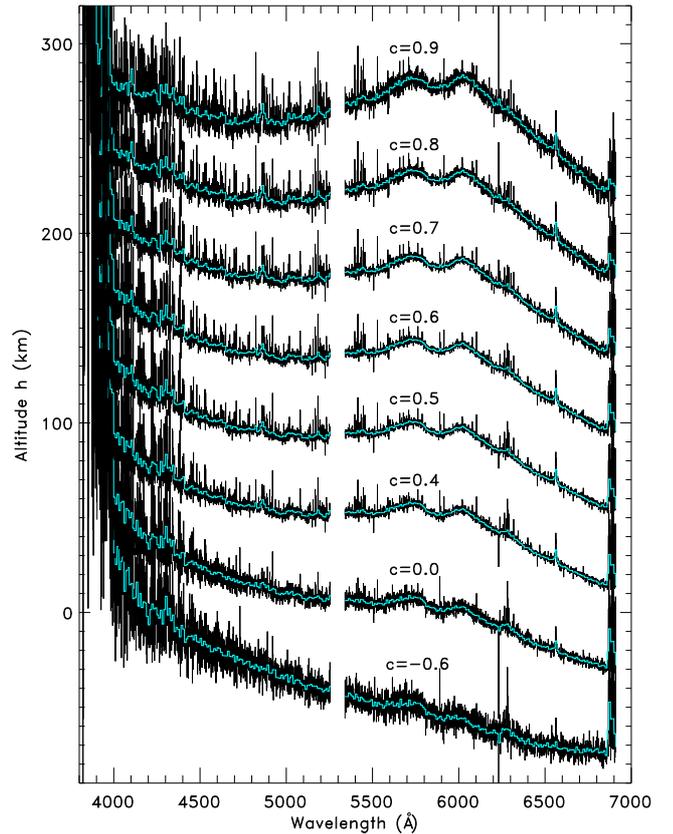}
   \caption{Altitude profiles obtained with HARPS and method~3 for different positions across the umbra edge, given by parameter $c$  in Eq.~\ref{eq:re} and shown in Fig.~\ref{best_re_points}. The vertical axis is the measured altitude, set to 23.8~km at at [4520-4540]\AA\ for the profile calculated for $c=0$. Other profiles are shifted vertically for clarity by steps of 40~km. The profiles in this figure are not built at HARPS native resolution, but at a smaller $\approx0.6$\AA/pixel resolution. The light-blue lines are 10\AA\ bins.}
    \label{best_d_profiles}
\end{figure}

\begin{figure}[h]
   \centering
   \includegraphics[width=9cm]{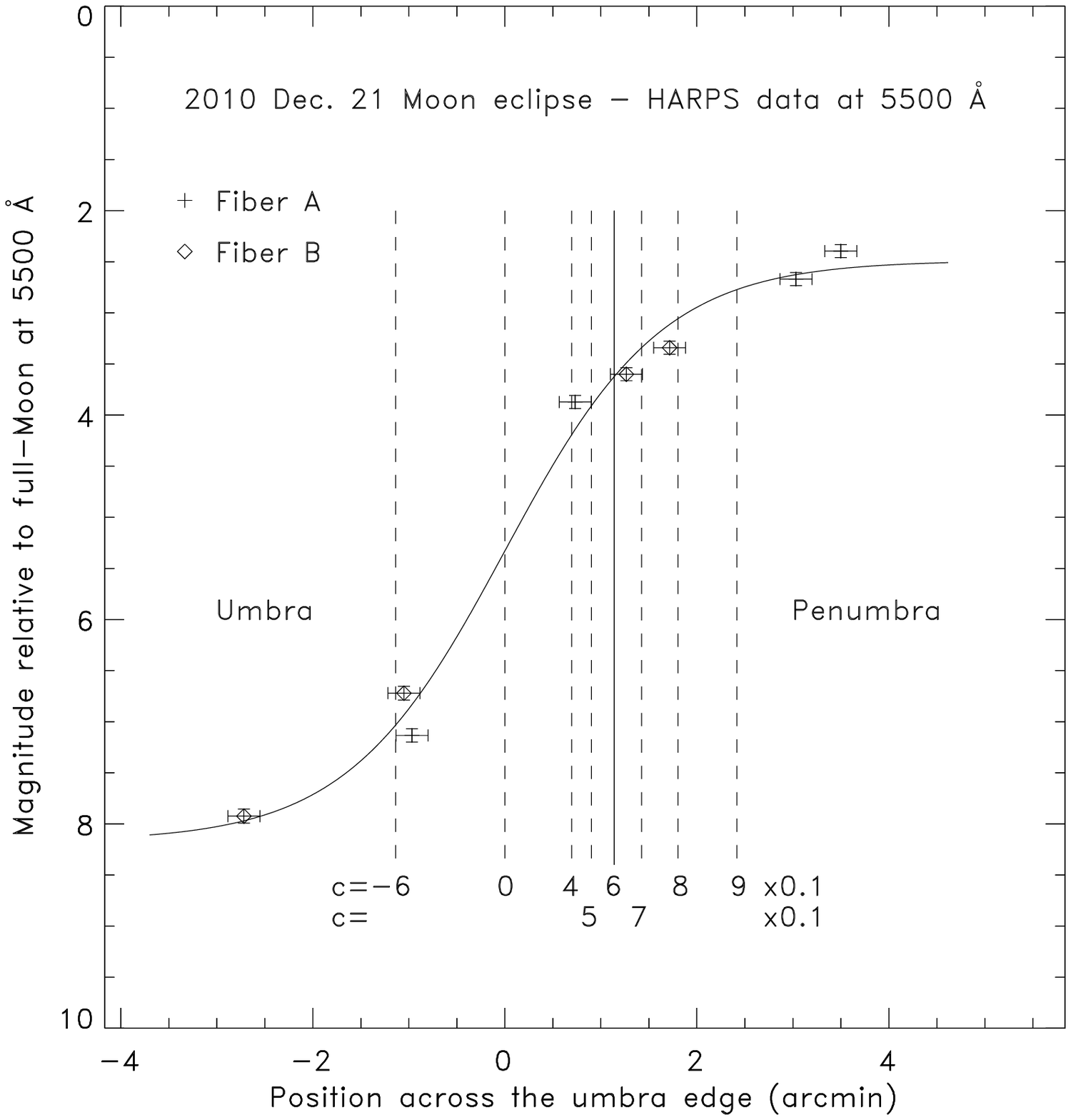}
   \caption{Positions across the umbra edge used to build the altitude profiles shown in Fig.~\ref{best_d_profiles}. The positions are defined by parameter $c$ (Eq.~\ref{eq:re}) and ranging from $c=-0.6$ to 0.9. The definition for the umbra edge is taken for $c=0.6$ (vertical solid line), at $\approx1$~arcmin in the penumbra side with respect to the inflection point, here fixed at abscissa=0. The umbra edge is only about one magnitude fainter than the penumbra at 4~arcmin from the inflection point.}
    \label{best_re_points}
\end{figure}

As explained in Sect.~\ref{method3}, the inflection point abscissa $a_2$ of the fit is not the best estimate of $r_e$ (in our context of Earth transit analysis, at least).  To explore the impact of $r_e$ on $h(\lambda)$, we write $r_e$ such that 
\begin{equation}
 \zeta(r_e)= a_0+c\times a_1.
  \label{eq:re}
\end{equation}
For the inflection point, the coefficient $c$ in front of half-amplitude $a_1$ of the $\tanh$ function is $c=0$ and $\zeta(a2)= a_0$.
Positive values of $c$ lead to positions on the penumbra side of the inflection point.  Fig.~\ref{best_d_profiles} shows a set of altitude profiles built for different positions across the umbra edge, with $-0.6 < c < 0.9$ as shown in Fig.~\ref{best_re_points}. The profiles versus $c$ change substantially: for $c=-0.6$, the profile is built from data on the umbra side of the inflection point, and the bump from the ozone Chappuis band is almost invisible. The mean negative slope of the profile also underlines the effect of refraction that shrinks the Earth umbra at redder wavelengths. This effect decreases when $c$ increases to positive values, with the profiles built from data on the penumbra side of the inflection point. The broad Chappuis band is detected. We finally define the umbra radius such as $\zeta(r_e)= a_0+0.6\ \times a_1$ (c=0.6). For this value, the noise in the altitude profile is close to minimum, and the bias from shallow umbra (that is, the mean negative slope of $h(\lambda)$) is reduced. With this definition, Fig.~\ref{best_re_points} puts the umbra edge at about 1~arcmin in the penumbra side with respect to the inflection point, and about one magnitude fainter than the penumbra at 4~arcmin from the inflection point. This choice is consistent with our reasoning developed Sect.~\ref{method3}.

\begin{figure*}[h]
   \centering
   \includegraphics[width=18cm]{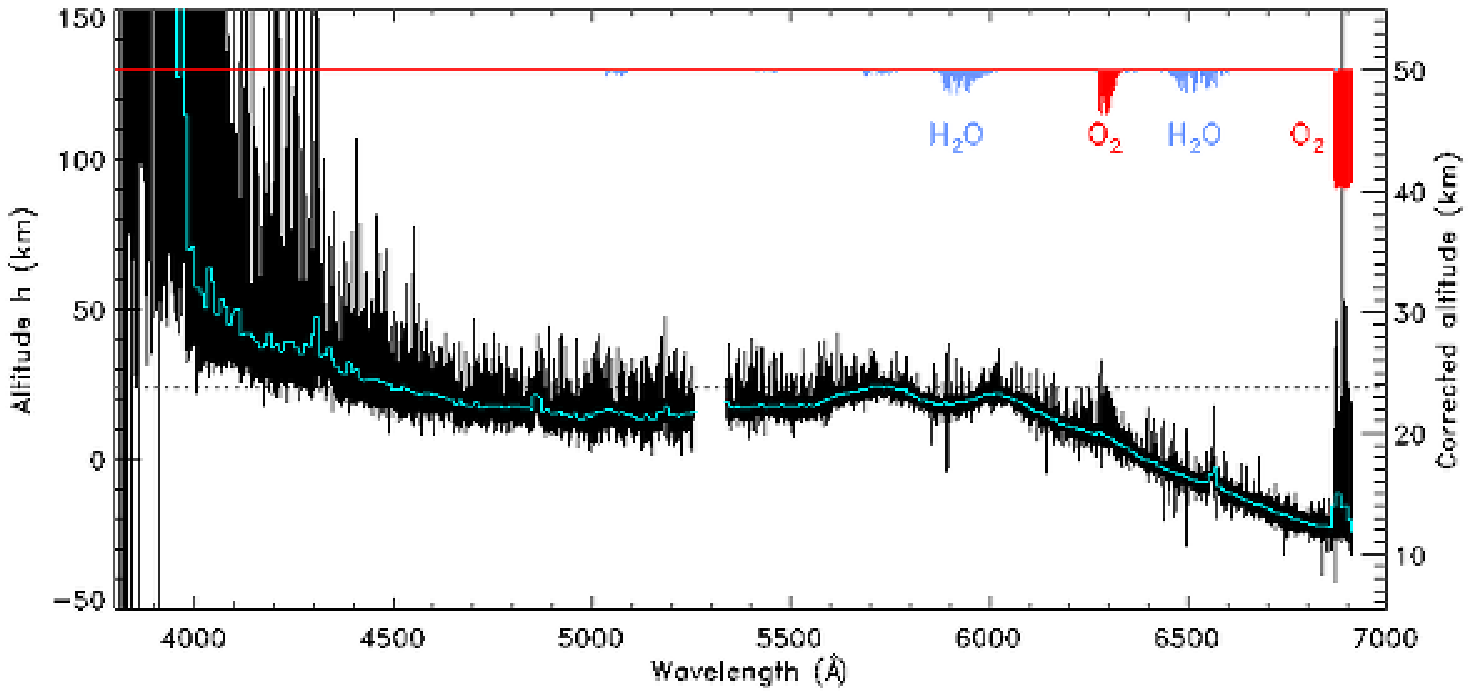}
   \caption{Altitude profile obtained with HARPS and method~3. The left vertical axis is the measured altitude, set to 23.8~km at at [4520-4540]\AA. The right vertical axis is the corrected altitude, if one considers the lowest region of $h$ close to altitude 12~km as in Fig.~\ref{HARPS_full_final}  and again $h=23.8$~km at [4520-4540]\AA. The horizontal dotted line represents the altitude 23.8~km.The light-blue line is a 10\AA\ bin superimposed on the full-resolution profile. Oxygen and water-vapour absorption spectra at arbitrary scale are shown.}
    \label{fig:HARPS_UPP}
\end{figure*}

\begin{figure}[h]
   \centering
   \includegraphics[width=9cm]{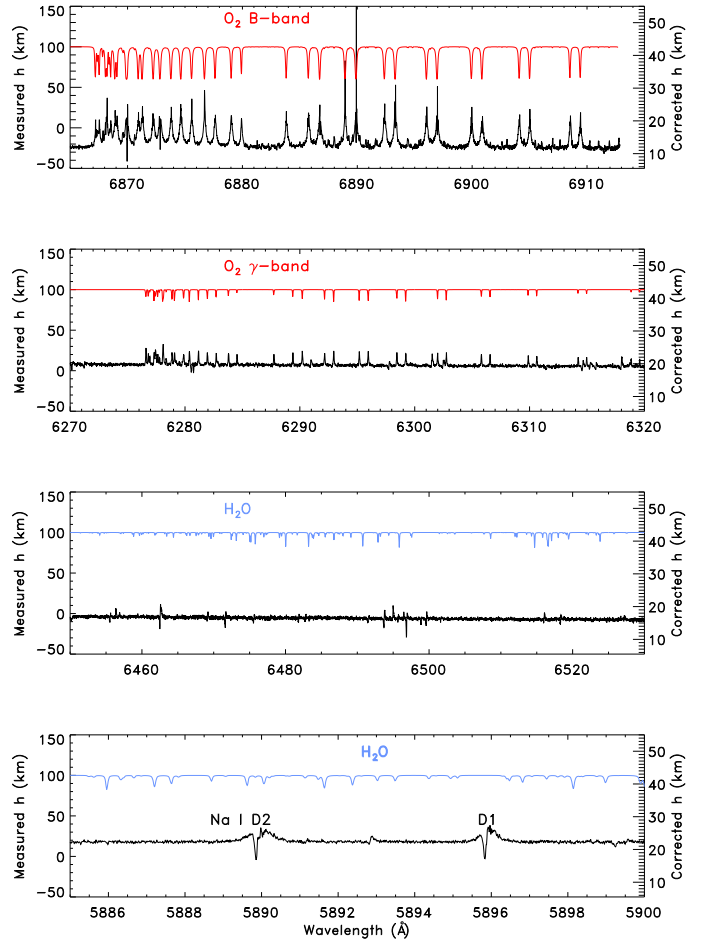}
   \caption{Zooms on the HARPS altitude profile from method~3 shown in Fig.~\ref{fig:HARPS_UPP}. The left vertical axis is the measured altitude, set to 23.8 km at [4520-4540]\AA. The right vertical axis is the corrected altitude, if one considers the lowest region of $h$ close to altitude 12~km as in Fig.~\ref{HARPS_full_final}   and again $h=23.8$~km at [4520-4540]\AA. Oxygen and water-vapour absorption spectra at arbitrary scale are shown. Water vapour is not detected above 16~km, suggesting that the amount of water vapour is too low to absorb enough to be visible at this altitude. The sodium layer Na I is not detected.}
    \label{fig:HARPS_UPP_zoom}
\end{figure}

The result with HARPS and method~3 is shown in Fig.~\ref{fig:HARPS_UPP} and Fig.~\ref{fig:HARPS_UPP_zoom}. Wide and narrow features of the Earth atmosphere are visible (Rayleigh, ozone Chappuis band, oxygen  $\gamma$ and B- bands). Water vapour is not detected, suggesting it is not present above 12~km. Compared with the profile obtained with method~1 (Fig.~\ref{HARPS_full_final}), the profile with method~3 shows a stronger negative slope from blue to red, leading to an altitude range about $\approx3.3$ times larger than with method~1. We attribute this effect to the fact that method~3 makes use of umbra spectra. The light in the umbra is reddened by Rayleigh scattering in the deep Earth atmospheric layers and refracted towards the umbra center. Therefore the umbra edge  in red light is more refracted towards the umbra center than at blue wavelengths. The umbra radius (Earth radius on the Moon) is thus smaller in the red than in the blue, due to refraction and Rayleigh scattering. This effect amplifies the measured altitude range across the full visible spectrum. To correct for this effect, the right vertical axis of Fig.~\ref{fig:HARPS_UPP} is calibrated with the method~1 result, by setting the altitude for two different points,that is, the lowest region of $h$ at altitude 12~km  and $h=23.8$~km at [4520-4540]\AA.

\begin{figure}[h]
   \centering
   \includegraphics[width=9cm]{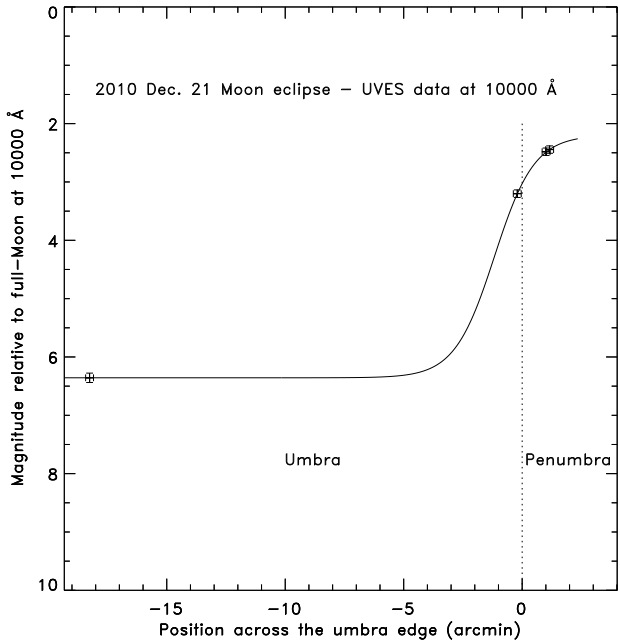}
   \caption{Measured magnitude across the umbra edge with UVES at 10000\AA.  The horizontal error bars are 20~arcsec long and represent the typical pointing error of the telescope.   Vertically, the $2\sigma$ error bars are the magnitude errors given by the UVES pipeline and the evaluated error in the local Moon albedo. The solid line is the model from Eq.\ref{eq:model} fitted through the data with parameter $a_3$ set to the mean value of $a_3$ found for the HARPS data. The vertical dotted line marks the edge of the umbra following our definition.}
    \label{fig:umbrapenumbraUVES}
\end{figure}

\begin{figure*}[h]
   \centering
   \includegraphics[width=18cm]{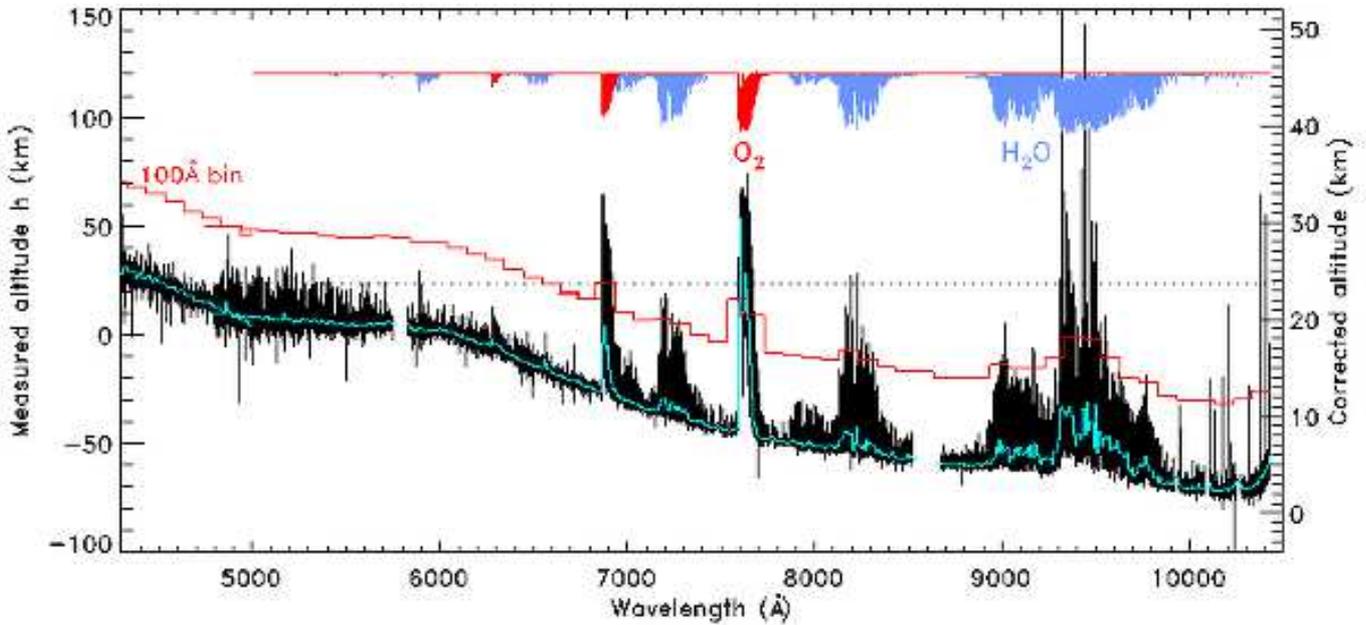}
   \caption{Altitude profile obtained with UVES and method~3. The left vertical axis is the measured altitude, set to 23.8~km at [4520-4540]\AA. The right vertical axis is the corrected altitude, if one considers the lowest region of $h$ close to altitude 0~km and again $h=23.8$~km at [4520-4540]\AA. The horizontal dotted line represents the altitude 23.8~km. A 10\AA\ bin is superimposed in light-blue on the full-resolution profile, and a 100\AA\ bin in red is also shown, shifted upwards for clarity. The profile for $\lambda \le \approx 4300$\AA\  is not shown, because the very weak umbra signal induces noise in the altitude profile which becomes noisy and $h$ unexpectedly decreases. Oxygen and water-vapour absorption spectra at arbitrary scale are shown.}
    \label{fig:UVES_UPP}
\end{figure*}

We process the UVES data in the same way. Nevertheless, since we have less observations points with UVES than with HARPS, it is not possible to keep four free parameters for the fitting function, the sigmoid $\zeta$. We have  to fix the $a_3$ parameter to the mean value obtained with HARPS. The parameter $a_3$ describes the steepness of the fit near the inflection point and is quite stable for HARPS at visible wavelengths: in the [6500-6900]\AA\ range, its average value is $\overline{a_3}=9.88\times10^{-3}$ with an RMS of $\sigma=2.2\times10^{-4}$. We fix $a_3$ to that mean value for UVES. An example of the fit through the UVES observations at 10000\AA\  is shown in Fig.~\ref{fig:umbrapenumbraUVES}.
Fig.~\ref{fig:UVES_UPP} shows the altitude profile obtained with UVES, showing the same features as for HARPS, with clear oxygen A and B bands, and water-vapour bands up to 10000\AA. As with the HARPS profile (Fig.~\ref{fig:HARPS_UPP}),  the profile obtained with UVES
displays a strong global negative slope. Here the profile is $\approx4.5$ times larger than with method~1. The right axis  in Fig.~\ref{fig:UVES_UPP} is calibrated considering that the lowest region of $h$ is close to altitude 0~km  and again that  $h=23.8$~km at [4520-4540]\AA. The factor of 4.5 is larger than with HARPS ($\approx3.3$) maybe because the slit positions with UVES are deeper in the umbra than with HARPS (see Tab.~\ref{tab:UVES_slit_pos}  and \ref{tab:HARPS_fiber_pos}).

In the blue ($\lambda \le \approx 4300$\AA), the signal in the umbra spectra is weak and the altitude profile becomes noisy and unexpectedly decreases (this part of the profile is not shown in Fig.\ref{fig:UVES_UPP}). The weak signal in the deepest lines of oxygen A-band and water vapour ($\approx9300-9500$\AA) in the umbra spectra is also responsible for spikes above 100~km at these wavelengths. To cancel out most of these spikes, we set a flux threshold for the umbra spectra of 1\% of the continuum below which the algorithm does not evaluate $h$.  At last, to have h=23.8~km at [4520-4540]\AA, we subtract 37~km from the DIC2 BLUE  profile. To have all parts of the profile connected in Fig.~\ref{fig:UVES_UPP}, we shift the DIC1 RED profile by -23~km and the DIC2 RED profile by +25~km.

\begin{figure}[h]
   \centering
   \includegraphics[width=9cm]{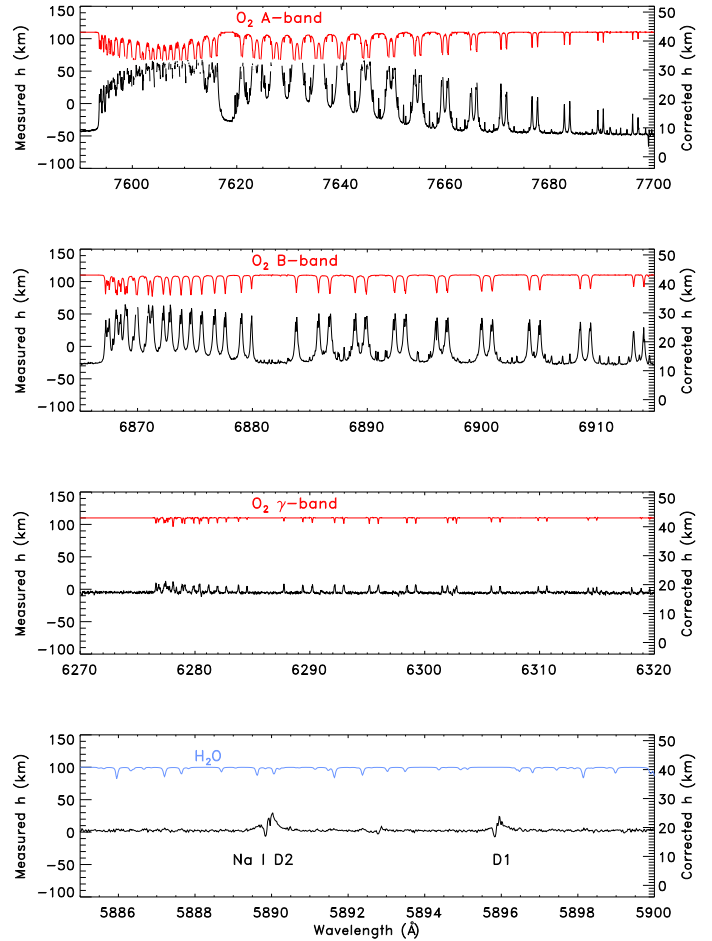}
   \caption{Zooms on the UVES altitude profile from method~3 shown in Fig.~\ref{fig:UVES_UPP}.  The UVES oxygen spectrum at arbitrary scale is shown. The three oxygen bands are clearly detected. Note that we were unable to calculate the core of the oxygen line for the A-band because the signal in the line core is too weak. The sodium layer is not detected. Water-vapour absorption is too weak in the Na I region to appear above 19~km.}
    \label{fig:ZoomO2_UVES_UPP}
\end{figure}

\begin{figure}[h]
   \centering
   \includegraphics[width=9cm]{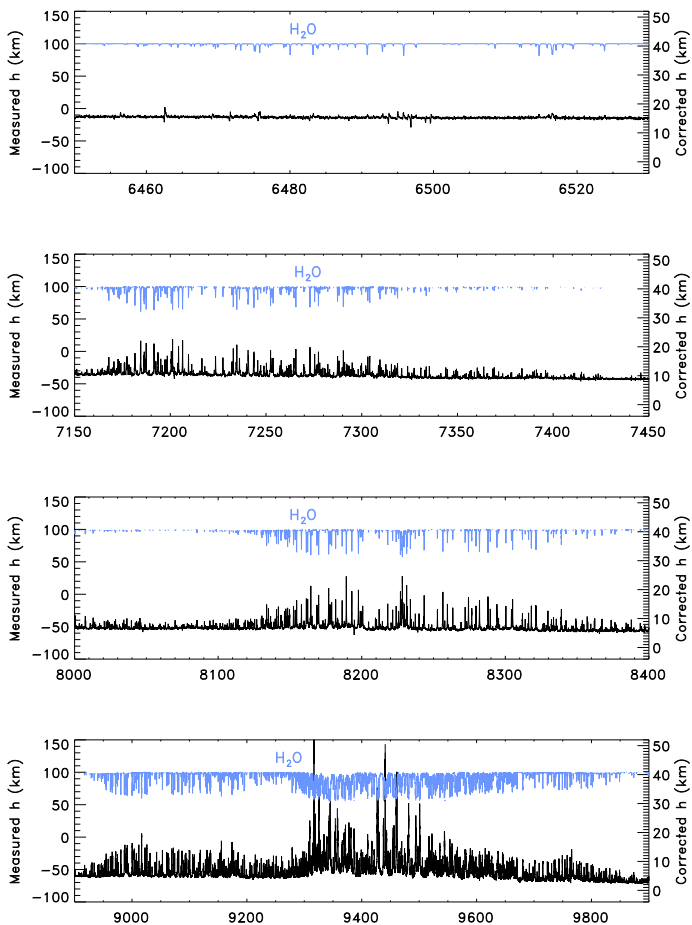}
   \caption{Zooms on the UVES altitude profile from method~3 shown in Fig.~\ref{fig:UVES_UPP}.  The UVES water-vapour absorption spectrum at arbitrary scale is shown.
   Water vapour is very well detected, even at an altitude above 20~km at $\lambda>8000$\AA\  where the water-vapour absorption is the strongest in the UVES wavelength domain. }
    \label{fig:ZoomH2O_UVES_UPP}
\end{figure}

The Fig.~\ref{fig:ZoomO2_UVES_UPP} and Fig.~\ref{fig:ZoomH2O_UVES_UPP} show zooms in the UVES full profile shown in Fig.~\ref{fig:UVES_UPP}. In Fig.~\ref{fig:ZoomO2_UVES_UPP}, the three oxygen bands are  clearly detected. Note that the core of the oxygen line for the A-band could not be calculated because the signal in the line core is too weak. The sodium layer is not detected, and no water vapour appears above 19~km near the sodium lines. At $\lambda>7000$\AA\ in Fig.~\ref{fig:ZoomH2O_UVES_UPP},  water vapour appears very well, even at an altitude higher than 20~km, because the water-vapour absorption bands become stronger in these longer wavelengths. Water vapour indeed presents a vertical distribution in the Earth atmosphere and is thus also present at higher altitudes but with lower volume densities. The altitude at which it is detected (where an optical thickness of about 1 is reached for a grazing line of sight) is of course higher in the stronger bands.

It is worth noting that the residuals of the Na I lines are very similar apart from the noise for HARPS and UVES (Fig.~\ref{fig:HARPS_UPP_zoom} and Fig.~\ref{fig:ZoomO2_UVES_UPP} respectively). This is expected because HARPS and UVES observed roughly the same solar crescents and thus experienced the same Doppler shift of the solar lines of the crescents that induced these residuals in the altitude profile.
Note also that no correction of the Ring effect is implemented here, either for HARPS or UVES. All spectra are indeed used in method~3, all at different depths in the penumbra and umbra. A simple empirical approach cannot reasonably be considered. A physical model of the Ring effect is much needed, taking into account the path length through the Earth atmosphere for the different positions observed in the umbra and penumbra.

\subsection{Possibilities with a future giant telescope}
\label{sect:EELTsimu}
We analyze how our results can be extrapolated to what a future ground-based giant telescope might be able to detect in the visible transmission spectrum of an Earth-like exoplanet observed in transit. 
We use a full model of the future E-ELT, a 39~m telescope. The focal instrument has the same throughput, pixel sampling, and detector characteristics as the EPICS instruments. The model takes into account the photon noise from the star, the thermal and readout detector noises, the zodiacal noise, and the absorption of the local atmosphere above the telescope (see \cite{hedelt2013} for a detailed description of the model).

We define a set of ten filters \textit{i)} to  sample the Chappuis band and the Rayleigh continuum and \textit{ii)} to isolate the most interesting narrow features in the profile shown in Fig.~\ref{fig:UVES_UPP}, that is, the oxygen A and B bands, and the water vapour around 9500\AA. The low spectral resolution 10\AA\  bin in this figure shows that the oxygen A-band peaks about 20~km above the lowest altitude around 10200\AA.  The Earth radius is thus 0.3\% larger and the transit 0.6\% deeper through a filter that isolates the oxygen A-band.  The Earth transit depth is $8.3\times10^{-5}$, and the transit variation at that spectral resolution for the $O_2$ band represents $7.5\times10^{-7}$ of the stellar flux.

According to Porto de Mello et al. (2006), there are 33 solar-type stars within 10~pc, 13 of which are reasonably similar to the Sun in terms mass, age, metallicity, and evolutionary status. Considering that the transit probability of  an Earth twin in front of its star is about 1\%, we understand that the probability of detection is low, even if all stars have a planet in their habitable zone. The Darwin All Sky Star Catalogue (\cite{kaltenegger2010}) lists 878 single FGK stars within 30~pc, a number fully consistent with the 33 solar-type stars numbered by Porto de Mello et al. (2006) for a 27 times smaller volume. Nevertheless, although the probability of transit detection significantly  increases with 878 targets, the flux for a star at 30~pc will be 9 times fainter than at 10~pc meaning that 9 times more transits will be necessary to reach the same S/N as for a target at 10~pc.

\begin{figure}[h]
   \centering
      \includegraphics[width=9cm]{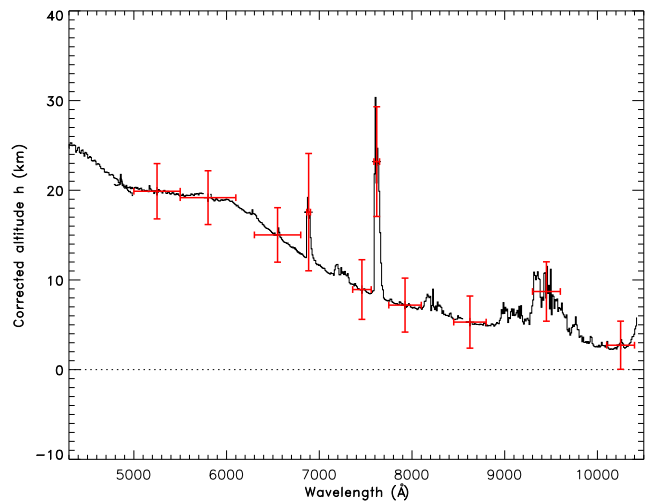}
   \caption{Simulated observations with the E-ELT of an Earth twin at 10 pc transiting a G2V star, at low spectral resolution and assuming no atmospheric perturbations (Tab.~\ref{tab:EELT-simu}). The simulated observations are shown in red. The horizontal bars represent the filter bandwidth and the vertical error bars are the $\pm 1\sigma$ estimated error from the model. The solid line is the 10\AA\ bin UVES profile (method~3).}
    \label{fig:EELTsimu}
\end{figure}

We consider an earth at 10~pc transiting a G2V star observed during a full night of 8~h (an Earth-like 13~h single transit being longer than the night). The simulation assumes that the atmosphere above the telescope is perfectly stable. The simulation results are given in Tab.~\ref{tab:EELT-simu} and shown in Fig.~\ref{fig:EELTsimu}. The result shows that transit photometry through the defined filters allows in principle the detection of the O$_2$ A-band with a detection level of $\approx2.3~\sigma$. The oxygen B and water-vapour bands will be more difficult to detect and will require more than one transit to reach a detection level $>$2$\sigma$. This is also true for the ozone Chappuis band, which is blended with the Rayleigh scattering.  

It is encouraging to see that the E-ELT will beat the photon noise in only one single transit to identify oxygen in the visible range on a terrestrial planet transiting a solar-type star at 10~pc.  But it will be extremely difficult, at least at low spectral resolution, to detect the $7.5\times10^{-7}$ flux variation in the $O_2$ band through the slightly fluctuating atmosphere above the telescope.

\begin{table}[h]
  \caption{Transit of an Earth twin at 10~pc observed for 8 hours by the E-ELT, at low spectral resolution and assuming no atmospheric perturbations. The S/N is calculated through ten different filters. The last filter 'Window' is for the atmospheric window. The filter widths range from 40\AA\ for the O$_2$ B to 600\AA\ for one of the filter sampling the Chappuis band. The values of $h$ are taken from the 10\AA\  bin of the UVES profile (Fig.~\ref{fig:UVES_UPP}).  The value $\delta$h is the altitude difference to the atmospheric window measured by the E-ELT.}
  \label{tab:EELT-simu}
  \begin{center}
    \leavevmode
    \begin{tabular}{lllll} \hline \hline              
Filter                  &  $\lambda$ range        &h        &    S/N        &  $\delta$h $\pm$ 1-$\sigma$ \\ 
	                  &             (\AA)                 & (km) &                   &      error (km)                   \\         \hline 
O$_3$ + Rayl. 1      &  5000-5500          & 19.9      & 11.3            &      17.2 $\pm$3.0                   \\
O$_3$ + Rayl. 2      &  5500-6100          & 19.2      & 12.3            &       16.5$\pm$2.9                   \\
O$_3$ + Rayl. 3      &  6300-6800           & 15.0      & 8.6            &       12.3$\pm$2.9                   \\
O$_2$-B                 &  6865-6905              & 17.6      & 2.5           &       14.8$\pm$6.4    \\
Rayl. 4                    &  7360-7560              &  8.9       &    3.1        &      6.2$\pm$3.2                   \\
O$_2$-A                 & 7590-7650             &  23.2     & 3.7          &        20.5$\pm$5.9   \\
Rayl. 5                    &  7750-8100            &     7.2   &    3.3         &          4.5$\pm$2.9                   \\
Rayl. 6                    &  8450-8800            &     5.3   &    2.3         &          2.6$\pm$2.8                 \\
Water  vapour       & 9300-9600              & 8.7      & 3.1               &        6.0$\pm$3.2  \\
Window           & 10100-10400               & 2.7        & 1.0           &      0$\pm$2.6           	 \\
\hline
    \end{tabular}
  \end{center}
\end{table}

We considered broadband and narrow-band photometry, with a spectral resolution of up to  $\approx170$. But it is worth noting that at a much higher spectral resolution (of the order of $\approx10^5$), the telluric lines become distinct from those of an exoplanet as soon as its radial velocity is sufficiently different from the BERV. Under these conditions, ground-based detection at high spectral resolution of O$_2$ or H$_2$O in an exoplanet atmosphere becomes easier, especially by cross-correlating the transit spectrum and an O$_2$ or H$_2$O spectrum, respectively (\cite{vidal-madjar2010}; \cite{snellen2013}).

\section{Summary and conclusion}
We have described the analysis of the observations of a lunar eclipse used as  a proxy to observe Earth in transit.
The results fully confirm those obtained with SOPHIE and the observation of the August 2008 eclipse  (\cite{vidal-madjar2010}) and extend the observed $h(\lambda)$ profile towards the near infrared thanks to the UVES data. They agree well with the models (\cite{ehrenreich2006}; \cite{kaltenegger2009}; \cite{snellen2013}; \cite{betremieux2013}). The Rayleigh increase in the blue is visible. Biogenic oxygen and ozone are also well visible in the atmosphere altitude profiles, and are therefore relevant detectable biosignatures to look for during transits of Earth-like exoplanets in the [4000-10500]\AA\ range. The oxygen A and B bands show up clearly with HARPS or UVES. 

Water vapour is well detected at $\lambda>$~8000\AA\  with UVES because water-vapour absorption is stronger at these wavelengths. Therefore even low densities of water vapour at these altitudes are detected. Water vapour is seen at an altitude of up to 20~km around 9500\AA. Moreover, Rayleigh scattering and ozone absorption are essentially absent  above 8000\AA\, therefore we are also able to probe deeper layers in the Earth atmosphere where low-altitude water vapour is also present at higher densities.   In HARPS data at $\lambda<$~7000\AA\, water vapour appears above 25~km with method~1 but remains below 25~km with method~2 and below 12~km with method~3, because water-vapour absorption is weaker in the visible range.

The results from UVES are nevertheless less reliable than those of HARPS, because the PWV was not stable during the beginning of the night when the calibration spectra were recorded on the full-Moon before the eclipse, which compromised the quality of the eclipse observation. Although we were able to recover the water-vapour content in the calibration spectra \textit{a posteriori}, the results for the detection of water vapour in the altitude profiles with UVES are not coherent between method~1, 2 and 3. This underlines the difficulties of these observations which were made through the local atmosphere above the telescope, while our goal was to measure atmospheric features at a planetary scale. Lunar eclipse observations done from space, either at low or high spectral resolution, would not suffer from the telluric pollution, with the additional benefit of an access to UV and the prominent Hartley band of ozone. 

Another bias in the recovering of the Earth atmosphere profile from a lunar eclipse is the effect of refraction in method~3 that makes use of refracted light from the umbra. We have shown that this effect can be partially corrected with the results from method~1, but the profiles - either from HARPS or UVES - still show a residual negative slope with respect to the profile from method~1 or 2.  

Finally these results provide input for the preparation of future observations with the next generation of giant telescopes. 
We showed that the E-ELT observing in the visible range at low spectral resolution and in the absence of atmospheric perturbations is in principle able to detect the O$_2$ A-band in the atmosphere of a transiting Earth twin. It will be extremely challenging in practice through the atmosphere. If we overcome this difficulty, especially with high spectral resolution spectroscopy, the O$_2$ signature may then be increasing from year to year (from transit to transit), as will  the enthusiasm of the observers.

\begin{acknowledgements}
      We thank Gaspare Lo Curto (HARPS), and Claudio Melo and Lorenzo Monaco (UVES) for their help in the preparation of these relatively unusual observations, as well as the teams who operated the telescopes that night. We also thank I. Snellen, and the referee, W. A. Traub, for their constructive comments on the manuscript. R. R. Querel acknowledges funding from Conicyt through Fondecyt grant 3120150.
\end{acknowledgements}

\end{document}